\documentclass[traditabstract]{aa}

\usepackage[stable]{footmisc}
\usepackage{graphicx}
\usepackage[varg]{txfonts}
\usepackage{hyperref}
\usepackage{natbib,twoopt}
\usepackage{lscape}
\usepackage{hyperref} 
\usepackage{subcaption}
\usepackage{nicefrac}
\bibpunct{(}{)}{;}{a}{}{,} 

\makeatother\bibpunct[, ]{(}{)}{;}{a}{}{,}

\newcommand{\farc}{\hbox{$.\!\!^{\prime\prime}$}}

\newcommand{\kms}{$\rm{km\,s^{-1}}$}
\newcommand{\hb}{H$\beta$} 
\newcommand{\ha}{H$\alpha$}
\newcommand{\hg}{H$\gamma$} 
\newcommand{\hd}{H$\delta$} 
 
\newcommand{\hii}{\mbox{H\,{\sc ii}}}

\newcommand{\sii}{[\ion{S}{ii}]} 
 
\newcommand{\oii}{[\ion{O}{ii}]}
 
\newcommand{\oiii}{[\ion{O}{iii}]}

\newcommand{\nii}{[\ion{N}{ii}]} 
\newcommand{\Msun}{$M_\odot$}
\newcommand{\Msunyr}{$M_\odot\,\rm{yr}^{-1}$}

\begin{document}

\title{The supermassive black hole coincident with the luminous transient ASASSN-15lh\thanks{Based on observations at ESO, Program IDs: 097.D-1054, 297.B-5035, 099.D-0115}}

\titlerunning{MUSE: Supermassive black hole}
\authorrunning{T.~Kr\"uhler et al.}

\author{T.~Kr{\"u}hler\inst{1}, M.~Fraser\inst{2,3}, G.~Leloudas\inst{4,5}, S.~Schulze\inst{4}, N.~C.~Stone\inst{6},  S.~van~Velzen\inst{7}, R.~Amorin\inst{8,9}, J.~Hjorth\inst{5}, P.~G.~Jonker\inst{10,11}, D.~A.~Kann\inst{12}, S.~Kim\inst{13,14}, H.~Kuncarayakti\inst{15,16}, A.~Mehner\inst{17}, A.~Nicuesa~Guelbenzu\inst{18}
}

\institute{Max-Planck-Institut f{\"u}r extraterrestrische Physik, Gie{\ss}enbachstra{\ss}e, 85748 Garching, Germany
\and School of Physics, O'Brien Centre for Science North, University College Dublin, Belfield, Dublin 4, Ireland.
\and Institute of Astronomy, University of Cambridge, Madingley Road, Cambridge CB3 0HA, UK 
\and Department of Particle Physics and Astrophysics, Weizmann Institute of Science, Rehovot 7610001, Israel  f
\and Dark Cosmology Centre, Niels Bohr Institute, University of Copenhagen, Juliane Maries Vej 30, 2100 K\o benhavn \O, Denmark 
\and Columbia Astrophysics Laboratory, Columbia University, New York, NY, 10027, USA
\and Department of Physics \& Astronomy, The Johns Hopkins University, Baltimore, MD 21218, USA
\and Cavendish Laboratory, University of Cambridge, 19 JJ Thomson Avenue, Cambridge CB3 0HE, United Kingdom
\and Kavli Institute for Cosmology, University of Cambridge, Madingley Road, CB3 0HA, United Kingdom
\and SRON, Netherlands Institute for Space Research, Sorbonnelaan 2, 3584 CA, Utrecht, The Netherlands
\and Department of Astrophysics/IMAPP, Radboud University Nijmegen, P.O. Box 9010, 6500 GL Nijmegen, The Netherlands
\and Instituto de Astrof\'{\i}sica de Andaluc\'{\i}a (IAA-CSIC), Glorieta de la Astronom\'{\i}a s/n, E-18008, Granada, Spain
\and Instituto de Astrof\'isica, Facultad de F\'isica, Pontificia Universidad Cat\'olica de Chile, Vicu\~{n}a Mackenna 4860, 7820436 Macul, Santiago, Chile
\and Max-Planck-Institut f\"{u}r Astronomie, K\"onigstuhl 17, 69117 Heidelberg, Germany
\and Finnish Centre for Astronomy with ESO (FINCA), University of Turku, V\"ais\"al\"antie 20, 21500 Piikki\"o, Finland
\and Tuorla Observatory, Department of Physics and Astronomy, University of Turku, V\"ais\"al\"antie 20, 21500 Piikki\"o, Finland
\and European Southern Observatory, Alonso de C\'{o}rdova 3107, Vitacura, Casilla 19001, Santiago 19, Chile 
\and Th\"{u}ringer Landessternwarte Tautenburg, Sternwarte 5, 07778 Tautenburg, Germany
}

\abstract{The progenitors of astronomical transients are linked to a specific stellar population and galactic environment, and observing their host galaxies hence constrains the physical nature of the transient itself. Here, we use imaging from the Hubble Space Telescope, and spatially resolved, medium-resolution spectroscopy from the Very Large Telescope obtained with X-Shooter and MUSE to study the host of the very luminous transient ASASSN-15lh. The dominant stellar population at the transient site is old (around 1 to 2 Gyr) without signs of recent star formation. We also detect emission from ionized gas, originating from three different, time invariable, narrow components of collisionally excited metal and Balmer lines. The ratios of emission lines in the Baldwin-Phillips-Terlevich diagnostic diagram indicate that the ionization source is a weak active galactic nucleus with a black hole mass of $M_\bullet = 5_{-3}^{+8}\cdot10^{8} M_\sun$, derived through the $M_\bullet$-$\sigma$ relation. The narrow line components show spatial and velocity offsets on scales of 1~kpc and 500~\kms, respectively; these offsets are best explained by gas kinematics in the narrow-line region. The location of the central component, which we argue is also the position of the supermassive black hole, aligns with that of the transient within an uncertainty of 170~pc. Using this positional coincidence as well as other similarities with the hosts of tidal disruption events, we strengthen the argument that the transient emission observed as ASASSN-15lh is related to the disruption of a star around a supermassive black hole, most probably spinning with a Kerr parameter $a_\bullet\gtrsim0.5$.}

\keywords{stars: individual: ASASSN-15lh, galaxies: supermassive black holes}
\maketitle

\section{Introduction}
\label{sec:Intro}

One of the most remarkable astronomical transients reported in recent years was ASASSN-15lh at a redshift of $z=0.232$; this transient was first discovered by the All-Sky Automated Survey for Supernovae \citep[ASAS-SN;][]{2014ApJ...788...48S} on 2015 June 24 \citep{2015ATel.7642....1N}. ASASSN-15lh is characterized by an exceptional brightness at peak ($M_u\sim-23.5$~mag), relatively high blackbody temperatures over a long period (between $12\,000$~K and $25\,000$~K in the first 300 days), and a relatively fast variability timescale of a few tens of days. Its radiated energy ($E_{\rm rad}\sim2\times10^{52}$~erg within 300 days), rapid temporal evolution, a strong rebrightening at 120 days \citep{2016ApJ...828....3B}, and largely featureless optical spectra \citep{2016Sci...351..257D, 2016NatAs...1E...2L} make this transient hard to classify within the established scheme of transient phenomena, and ASASSN-15lh has prompted a number of theoretical models \citep[e.g.,][]{2015MNRAS.454.3311M, 2016ApJ...817L...8B,2016MNRAS.459L..21K,2017arXiv170504689C,2016ApJ...820L..38S}, which have attempted to explain its remarkable observational features.

ASASSN-15lh was initially suggested \citep[][]{2016Sci...351..257D, 2017MNRAS.466.1428G} to be a hydrogen-poor superluminous supernova (SLSN), an explosive phenomenon from the collapse of a massive star \citep{2011Natur.474..487Q}. Other authors, however, have subsequently disagreed with this classification \citep{2016NatAs...1E...2L,  2017ApJ...836...25M}. They argued that the observations are more consistent with those of thermal tidal disruption events \citep[TDEs; e.g.,][]{1999A&A...349L..45K, 1999A&A...343..775K, 2008ApJ...676..944G, 2011ApJ...741...73V, 2012Natur.485..217G, 2014ApJ...793...38A}, the luminous emission from the accretion of stellar material after the disruption of a star by a supermassive black hole (SMBH) at the center of a galaxy \citep[e.g.,][]{1988Natur.333..523R, 1989ApJ...346L..13E}. This possibility was initially considered unfeasible, primarily because the black hole mass $M_\bullet$ of the central black hole in the massive galaxy that hosted ASASSN-15lh seemed too high \citep{2015ATel.7776....1P}. Above the Hills mass $M_{\mathrm{Hills}}\sim10^{8}$~\Msun, the tidal radius of a solar-mass star lies within the minimum pericenter for parabolic orbits around a non-rotating black hole, and no electromagnetic radiation should escape these kinds of encounters \citep{1975Natur.254..295H}. However, the limits on the maximum black hole mass that can produce luminous emission are relaxed for spinning black holes \citep{2012PhRvD..85b4037K}, such that a TDE has remained a physically viable model to produce ASASSN-15lh \citep{2016NatAs...1E...2L}, possibly also explaining the relatively low volumetric rate of similar events \citep{2017arXiv170703458V}.

There are four fundamental shortcomings with SLSN-related models to explain the observations of ASASSN-15lh. The first problem is the absence of broad metal absorption lines that are typically observed in SLSN in their optical spectra. One of the most salient features in early SLSN spectra, for example, arises from an \ion{O}{ii}-doublet at 4100~\AA~and 4400~\AA~\citep{2011Natur.474..487Q}. An absorption line at the expected wavelength of the bluer \ion{O}{ii} transition is detected in early ASASSN-15lh spectra, however, this line must be also accompanied by the redder transition \citep{2016MNRAS.458.3455M}, which would have clearly been detected in the available data if it were present \citep{2016NatAs...1E...2L, 2017ApJ...836...25M}. Secondly, the UV absorption line spectrum of ASASSN-15lh is dominated by narrow high-ionization lines, such as \ion{N}{v} or \ion{O}{vi}, which is unusual for phenomena related to the collapse of massive stars. Thirdly, the temporal evolution of the transient would be unprecedented for SLSN, and its temperature and bolometric luminosity are inconsistent with previous observations of SLSN and the theoretical expectation of an expanding photosphere for these events. Finally, the galactic environment of ASASSN-15lh is in stark contrast to previous hydrogen-poor SLSN hosts, which have exclusively been strongly star forming and generally have low-stellar mass and metallicity \citep[e.g.,][]{2015MNRAS.449..917L, 2016ApJ...830...13P, 2016arXiv160504925C}. However, the extreme observational characteristics of ASASSN-15lh are clearly unprecedented not only for SLSN, but also among bona fide, optically selected TDEs.

The physical properties of galaxies observed to host TDEs are very different from those of SLSNe. Such galaxies exhibit a curious preference  \citep{2014ApJ...793...38A, 2016ApJ...818L..21F, 2017arXiv170702986G, 2017arXiv170701559L} for E+A galaxies \citep{1996ApJ...466..104Z}. These galaxies are dominated by the stellar light from A-type stars and show characteristically strong Balmer absorption. They are thought to be the result of a starburst several hundred million years in the past, possibly triggered by a galaxy merger \citep{1996ApJ...466..104Z}. In fact, spatially resolved spectroscopy of the nearby E+A galaxy hosting the TDE ASASSN-14li \citep{2016ApJ...830L..32P} reveals distinctive tidal tails in ionized gas emission, which is likely the result of a recent galaxy merger and ionization from an active galactic nucleus (AGN). The reason for the over-representation of E+A galaxies within the TDE host sample is an active subject of discussion; that is, physical processes such as stellar dynamics around binary black holes or increased stellar densities in the galaxy center have been discussed in recent literature \citep[e.g.,][]{2016ApJ...825L..14S, 2017ApJ...835..176F}.

Because of the strong differences in the physical properties of the host galaxies of SLSNe and TDEs and the constraints that the environment can imply for the physical nature of the transient, we obtained sensitive long-slit, medium-resolution and spatially resolved integral field unit (IFU) spectroscopy of the field of ASASSN-15lh. These observations are used to pinpoint the exact position of ASASSN-15lh within its host, reveal emission lines from ionized hydrogen and metal lines, constrain the mass of the central black hole, and probe the large-scale environment.

Throughout the paper, we adopt concordance cosmology with Planck parameters ($H_0=67.3\,\mathrm{km}\,\mathrm{s}^{-1}\,\mathrm{Mpc}^{-1}$, $\Omega_\mathrm{m}$=0.315, $\Omega_\Lambda$=0.685, \citealt{2014A&A...571A..16P}), a \citet{2003PASP..115..763C} initial mass function (IMF), and report errors at the 1~$\sigma$ confidence level.

\section{Observations and data analysis}
\label{sec:Obs}

\subsection{Hubble Space Telescope imaging}
\label{obs:hst}

\begin{figure*}
  \includegraphics[width=0.999\linewidth]{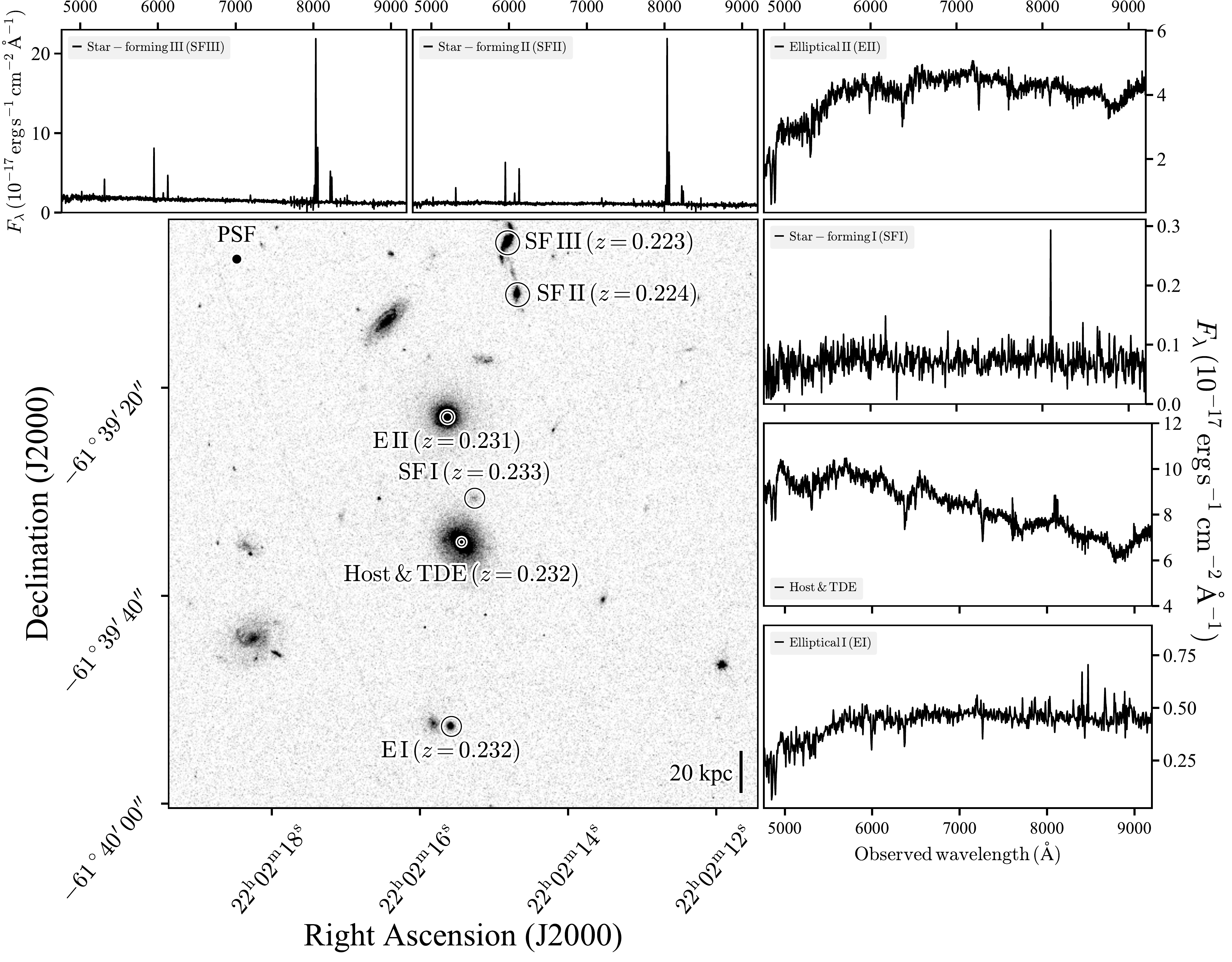}\caption{HST WFC3 image in the F606W filter in the \textit{main panel}. The image spans approximately 60" by 60", which is 220~kpc by 220~kpc at the redshift $z=0.232$ of ASASSN-15lh and corresponds to the field of view of our MUSE integral field spectroscopy (Section~\ref{obs:muse}). The six panels above and to the right of the WFC3 image show extracted MUSE spectra of the host (plus transient) and the five additional galaxies at a similar redshift. These galaxies are denoted SFI, SFII, and SFIII for the three star-forming galaxies and EI and EII for two passive ellipticals in the main image. Other galaxies visible in the image are either fore- or background sources. The size of the MUSE point spread function and a physical scale at $z=0.232$ are indicated in the top left\ and bottom right corner of the image, respectively.}
\label{fig:fc}
\end{figure*}

\begin{figure}
\begin{subfigure}{.2425\textwidth}
  \includegraphics[width=0.999\linewidth]{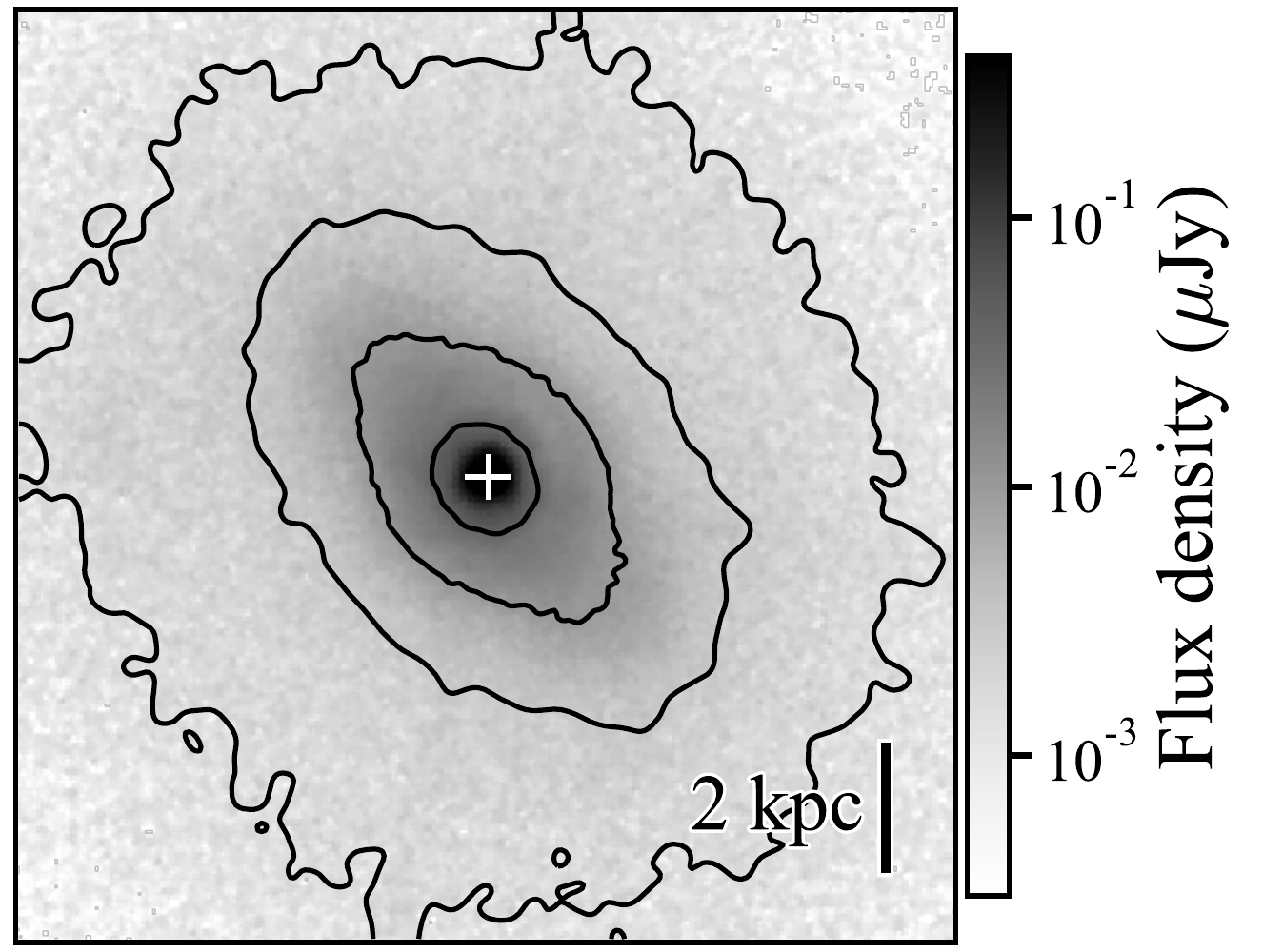}
\end{subfigure}
\begin{subfigure}{.2425\textwidth}
  \includegraphics[width=0.999\linewidth]{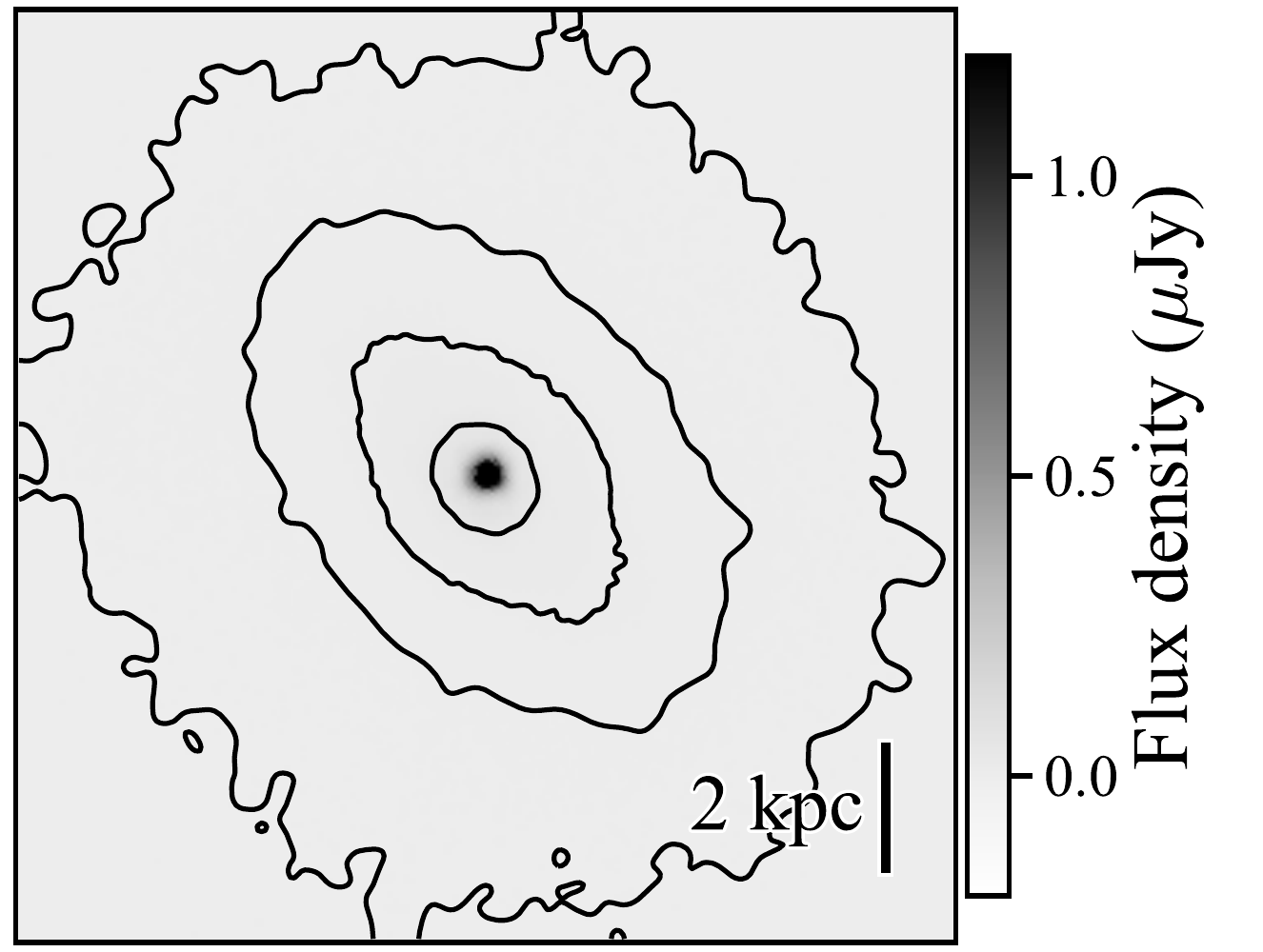}
\end{subfigure}
\caption{Zoom in on the HST WFC3/F606W image {to highlight the ASASSN-15lh host galaxy morphology (\textit{left panel}, logarithmic scaling) and the transient (\textit{right panel}, linear scaling). Both panels show the same image, sky area, and contours, but have different color bars.} The image size is approximately 4\farc{2} by 4\farc{2}, which is 16~kpc by 16~kpc at the redshift of ASASSN-15lh. North is up and east is to the left, and a physical scale at $z=0.232$ is indicated in the bottom right corner of the image, respectively. {The position of the transient is indicated with a white cross in the left panel.}}
\label{fig:fczoom}
\end{figure}

The field of ASASSN-15lh was observed with Hubble Space Telescope (HST) Wide-Field Camera 3 (WFC3) under program 14346 (PI: C. Kochanek). A total of six exposures of 416~s integration time each were obtained on 2016-08-11 (433 days after peak brightness on 2015-06-05; \citealt{2016Sci...351..257D}) in the F606W filter through Director's Discretionary Time and made public in the HST archive. We downloaded the processed and CTE-corrected data and drizzled them onto a single frame with a pixel scale of 0\farc{025}\,$\mathrm{px}^{-1}$ or 93\,pc\,px$^{-1}$ at the redshift of ASASSN-15lh. The transient is clearly detected at high significance as a bright point source with a full width at half maximum (FWHM) of 0\farc{07} above the continuum emission of the galaxy {(Fig.~\ref{fig:fczoom})}. Tying the WFC3 astrometry to 10 sources from the Gaia DR1 catalog \citep{2016A&A...595A...2G, 2016A&A...595A...1G}, we measured a position of $\mathrm{R.A.~(J2000)}=$22:02:15.4263, $\mathrm{Decl.~(J2000)} = -$61:39:34.910 in the astrometric reference frame defined by Gaia. The positional uncertainty is dominated by the root-mean-square difference to the astrometric tie objects, which is 8 mas in each coordinate (30 pc comoving). A cutout from this image with a size of 1\arcmin\,by 1\arcmin~and centered around the transient position is shown in the central panel of Figure~\ref{fig:fc}.

The host is elongated along the northeast to southwest direction (Figure~\ref{fig:fczoom}), and together with the lack of recent star formation (Section~\ref{sec:ionsource}) and broadband photometric colors, it displays the typical characteristics of early-type elliptical (E4 in this case) or lenticular (S0) galaxies \citep[e.g.,][]{2009ARA&A..47..159B}. Within hierarchical structure formation, these galaxies are often thought to be the result of mergers in particular in group environments where galaxy encounters are frequent \citep[e.g.,][]{2005A&A...437...69B, 2011MNRAS.415.1783B}. However, other physical processes such as ram pressure stripping or quenching of spirals due to starvation could play a role as well, such that the formation of in particular S0s remains the subject of active research.

\subsection{X-Shooter long-slit spectroscopy}
\label{obs:xs}

We initiated ground-based observations of ASASSN-15lh with X-Shooter \citep{2011A&A...536A.105V}, a cross-dispersed long-slit spectrograph mounted at ESO's Very Large Telescope (VLT) Unit Telescope (UT) 2. X-Shooter operates in three arms, simultaneously covering the wavelength range of $3000$~\AA\,to $25\,000$~\AA\, with a resolution between 30~\kms\, and 60~\kms\, depending on the slit width and arm. In total, we obtained approximately 5700~s of integration split over two nights (2016-07-02 and 2016-08-02, which is 393~days and 424~days after peak brightness) through program 297.B-5035 (PI: M.~Fraser). We used X-Shooter slit widths of 1\farc{0} (3000~\AA\,to 5500~\AA), 0\farc{9} (5500\,\AA~to 10\,000~\AA) and  0\farc{9} (10\,000~\AA\ to 25\,000~\AA), respectively, which were centered on the transient and oriented along the parallactic angle. Given that the transient aligns with the brightest part of the galaxy, these spectra are hence a superposition of transient and galaxy light.

The X-Shooter spectroscopy was reduced in a similar manner as described in detail in \citet{2015A&A...581A.125K}, making use of the ESO pipeline in its version \texttt{2.7.1} \citep{2006SPIE.6269E..80G, 2010SPIE.7737E..56M} and custom-written methods and tools. The data were flux-calibrated against the nightly spectrophotometric standard, which was LTT7987 on 2016-07-02 and EG274 on 2016-08-02, and extracted using variance weighting. The signal to noise of the final spectrum is between 20 and 30 per spectral bin of size 0.4~\AA~in the observed wavelength range between 3800~\AA~ and 9700~\AA, and somewhat lower above and below.

\subsection{MUSE integral-field spectroscopy}
\label{obs:muse}

We also used the Multi-Unit Spectroscopic Explorer (MUSE; \citealt{2010SPIE.7735E..08B}) at VLT UT4 to obtain spatially resolved spectroscopy of the field around ASASSN-15lh. The MUSE instrument is a state-of-the-art integral field unit (IFU) with an unprecedented combination of sensitivity, spatial sampling (spaxel size of 0\farc{2}\,x\,0\farc{2}), wavelength coverage (4750\,\AA\,to 9350\,\AA), and resolving power ($R=1500$ to $R=3000$ increasing from blue to red wavelengths). We obtained IFU spectroscopy for 3600~s  of total integration on 2016-08-26, or 448 days after peak, under program 097.D-1054 (PI: S.~Kim). The VLT/MUSE point spread function (PSF) of this epoch defines the effective spatial resolution and has a FWHM of 0\farc{8} at 8000~\AA.

A second epoch of MUSE spectroscopy was obtained on 2017-06-28 (754 days after peak) under ESO program 099.D-0115 (PI: T.~Kr\"uhler). A total of 2800~s of exposure time on source lead to a similar depth as in the earlier epoch, but with a somewhat worse PSF with a FWHM=1\farc{0} at 8000~\AA. Given the slightly better spatial resolution of the data set from 2016, most of the quantities, results, and plots were derived using the earlier MUSE spectroscopy. The observations from the second epoch yield fully consistent results.

Initial data processing was performed via the MUSE pipeline\footnote{http://www.eso.org/sci/software/pipelines/} (version \texttt{1.6.2}; \citealt{2014ASPC..485..451W}), which produces a fully reduced and sky-subtracted three-dimensional data cube that is calibrated in wavelength, flux, and the two astrometric dimensions. Starting with this pipeline-produced data cube, we used third-party software to correct for telluric absorption \citep[\texttt{molecfit};][]{2015A&A...576A..77S} and sky-line residuals  \citep[\texttt{zap};][]{2016MNRAS.458.3210S}, and our own software for the analysis.

We further corrected the MUSE flux scale through synthetic photometry of the star at $\mathrm{R.A.~(J2000)}=$22:02:11.92, $\mathrm{Decl.~(J2000)} = -$61:39:46.6 (Figure~\ref{fig:fc}) and the comparison to its $r$- and $i$-band magnitudes from \citet{2016NatAs...1E...2L}. To map the accurate, HST-derived position from Section~\ref{obs:hst} onto the MUSE data cube, we first reconstructed several images centered at various wavelengths from the MUSE integral field spectroscopy. The MUSE field of view contains a handful (5-7, depending on the wavelength range used to reconstruct the MUSE images) of comparison sources, which we can then used to register the MUSE astrometry against the HST imaging with a linear transformation\footnote{The optical distortion of MUSE, visible as a trapezoidal shape of the field of view, is already removed from the data cube through the astrometric correction applied within the MUSE pipeline.}. The registration process using different reconstructed images yields consistent results within a typical scatter smaller than 40~mas (or 0.2~MUSE pixels). This places ASASSN-15lh in our MUSE data cube with a total accuracy better than 150~pc in each coordinate.

\section{Results}
\label{sec:Res}

\subsection{Modeling the spectral continuum}
\label{sec:pPXF}

\begin{figure}
  \includegraphics[width=0.999\linewidth]{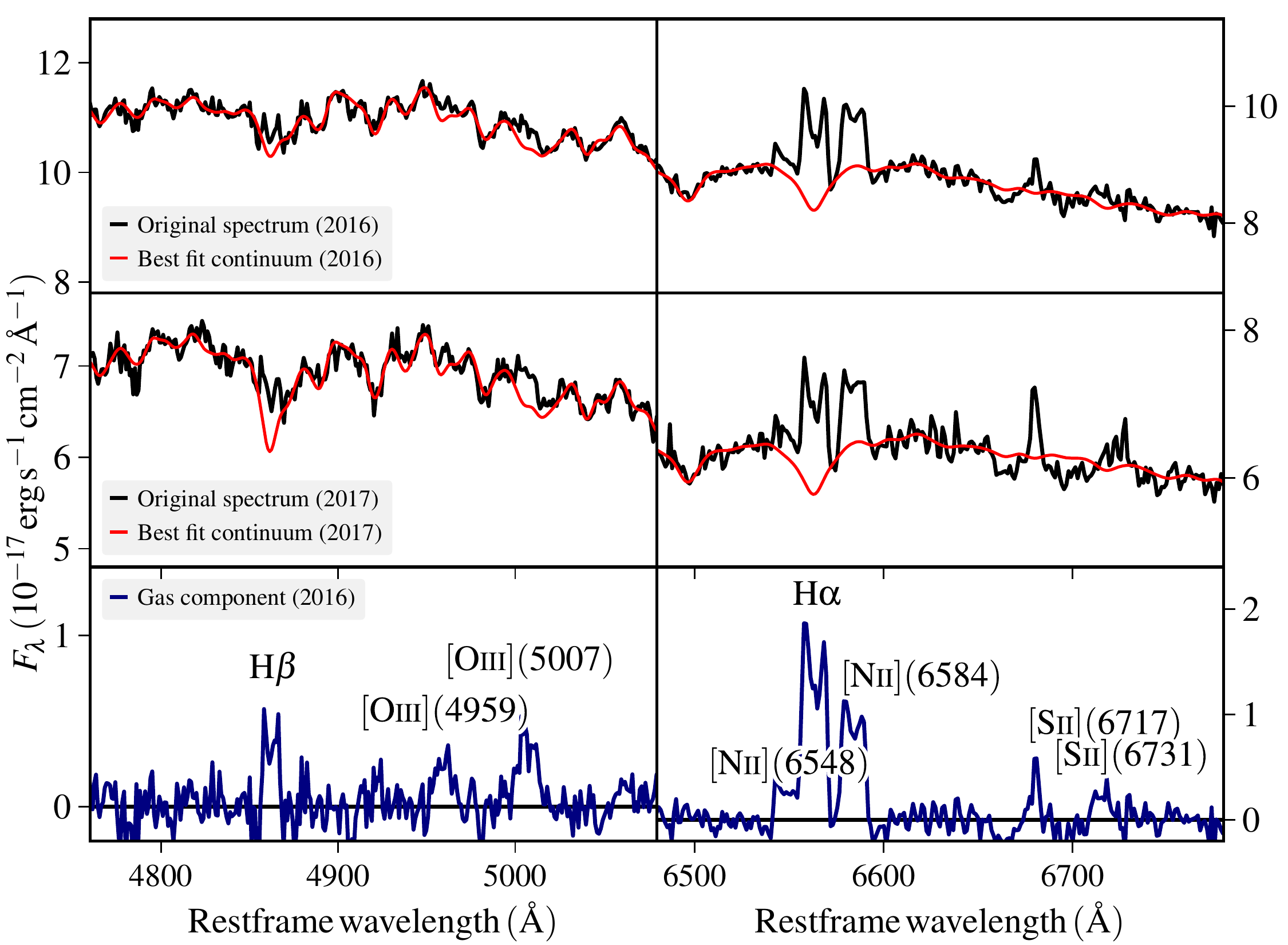}
\caption{Decomposition of continuum and gas emission. The \textit{upper four panels} show the observed spectrum in black and the best fit continuum from \texttt{pPXF} in red in the wavelength range of \hb~and \oiii~on the left, and \ha~and \nii~on the right side. The \textit{top} row is derived from the MUSE spectra in 2016 (424 days after peak), the \textit{middle} row shows the data from 2017 (754 days after peak). The \textit{lower two panels} show the emission-line contribution 
after subtraction of the continuum component for the 2016 data. A similar procedure applied to the 2017 data yields consistent results for the line fluxes and line ratios.}
\label{fig:stargas}
\end{figure}

Emission lines caused by the recombination of hydrogen and decay of collisionally excited states of metal ions provide information concerning the physical properties of the plasma and its ionization source. In particular the ratios between Balmer and metal lines are fundamental to ascertain whether star formation is present in the host. The Balmer lines, however, are a superposition of stellar absorption lines, emission lines from the ionized gas, and a potential contribution of the transient itself. 

We disentangled these various components by modeling the observed continuum with stellar templates, while emission from the transient is represented with a low-order polynomial. An illustrative example of this procedure using penalized pixel fitting \citep[\texttt{pPXF};][]{2004PASP..116..138C, 2017MNRAS.466..798C} is given in Fig.~\ref{fig:stargas}.
where we show the central component of the host extracted from the MUSE cube to highlight a couple of features. Firstly, there are obvious detections of multiple emission lines, which correspond to the transitions of \hb, \oiii($\lambda\lambda$4959, 5007), \nii($\lambda\lambda$6548, 6584), \ha\, and potentially \sii($\lambda\lambda$6717, 6731), even though the significance of the latter is not particularly convincing (Figure~\ref{fig:stargas}) and depends somewhat on the details of the subtraction and telluric correction.

The superposition of the \ha~and \nii~ complex is what was interpreted as a broad \ha\, emission in the lower resolution spectra of \citet{2016NatAs...1E...2L}. The detected lines are constant between the two MUSE epochs while the transient significantly declines over the time period of one year, and we hence confirm the line emission as coming from the host (and not from the transient). We thus corroborate the identifications in \citet{2017ApJ...836...25M} and their interpretation that the lines originate in the host galaxy. Secondly, each of these lines is obviously not well described with a single Gaussian line shape and shows strong velocity structure. We return to the line shape and discuss it in detail in Sect.~\ref{sec:prof}.

\subsection{Stellar kinematics and mass of the central black hole}

\begin{figure}
  \includegraphics[width=0.999\linewidth]{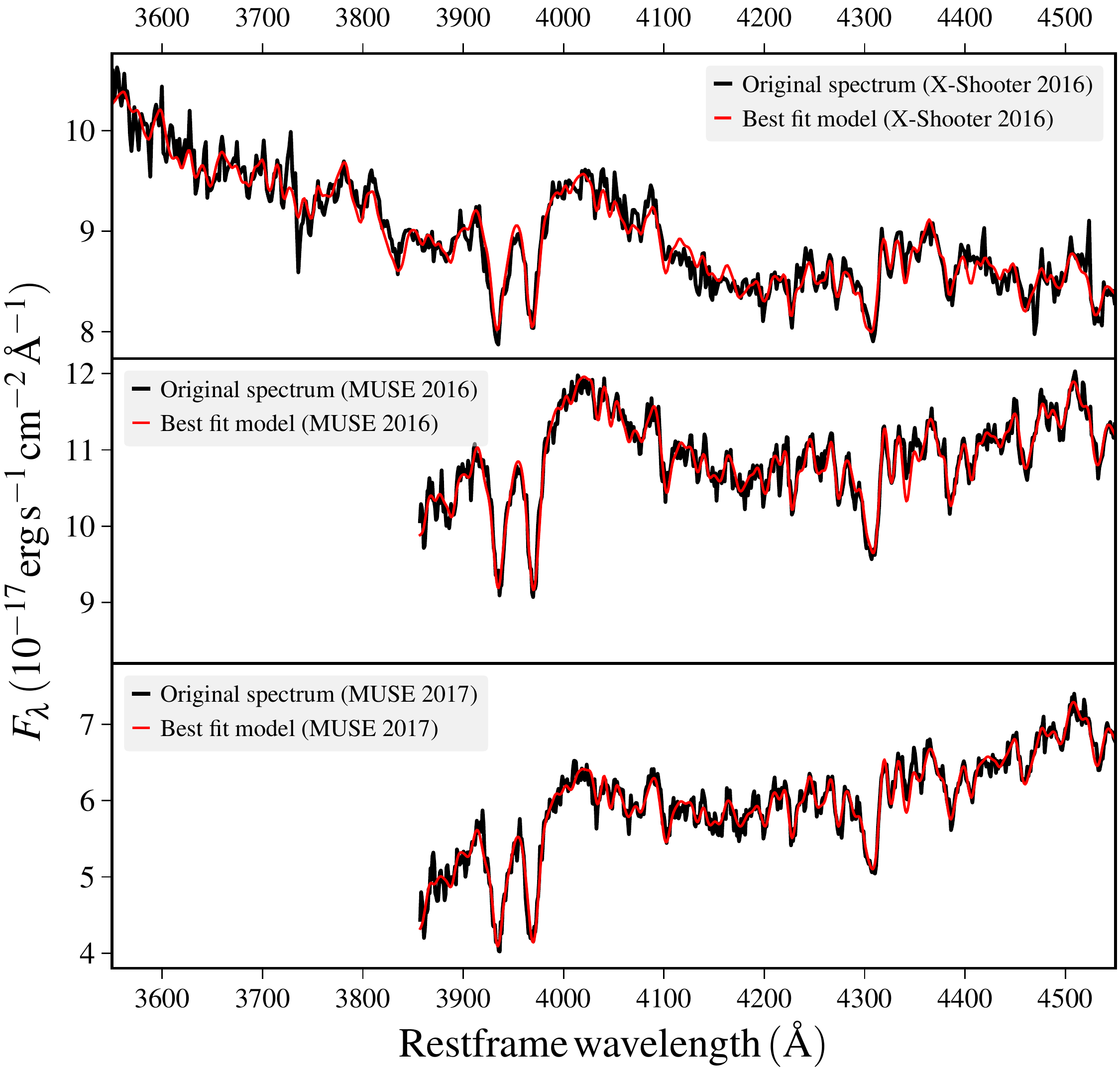}
\caption{Empirical model of transient plus stellar emission fit to the X-shooter {and MUSE} spectra for measuring the stellar kinematics in the wavelength range of the strong \ion{Ca}{II}~H+K doublet, \hd, \hg, and the G band. The emission from the transient is clearly evident through the decrease in flux between 2016 and 2017, and a continuum that is substantially bluer than what would be expected from the Gyr-old stellar population (Sect.~\ref{sec:spop}) of the host galaxy alone.}
\label{fig:stargas_sig}
\end{figure}

\begin{figure}
\begin{subfigure}{.2425\textwidth}
  \includegraphics[width=1.0\linewidth]{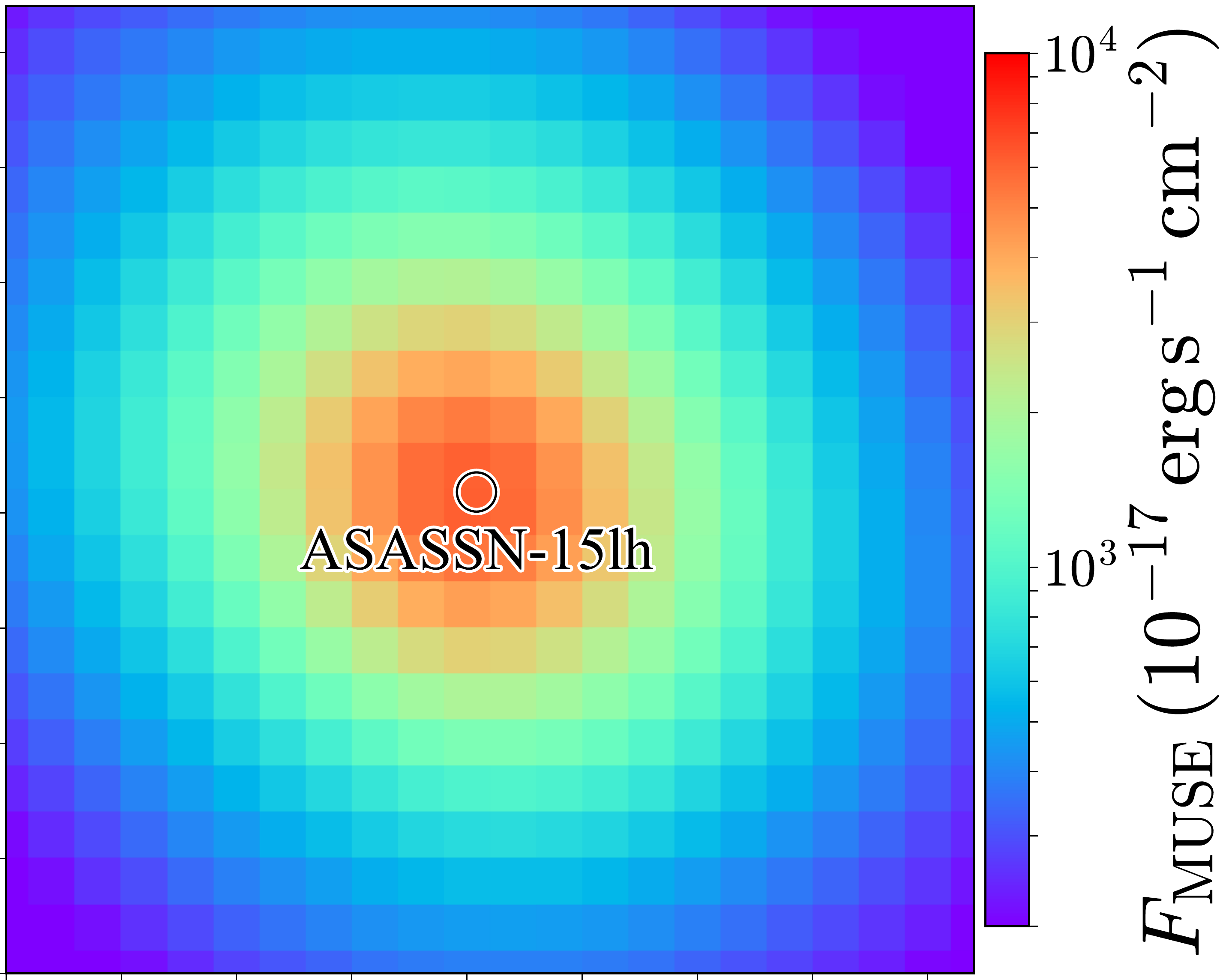}
\end{subfigure}
\begin{subfigure}{.2425\textwidth}
  \includegraphics[width=1.0\linewidth]{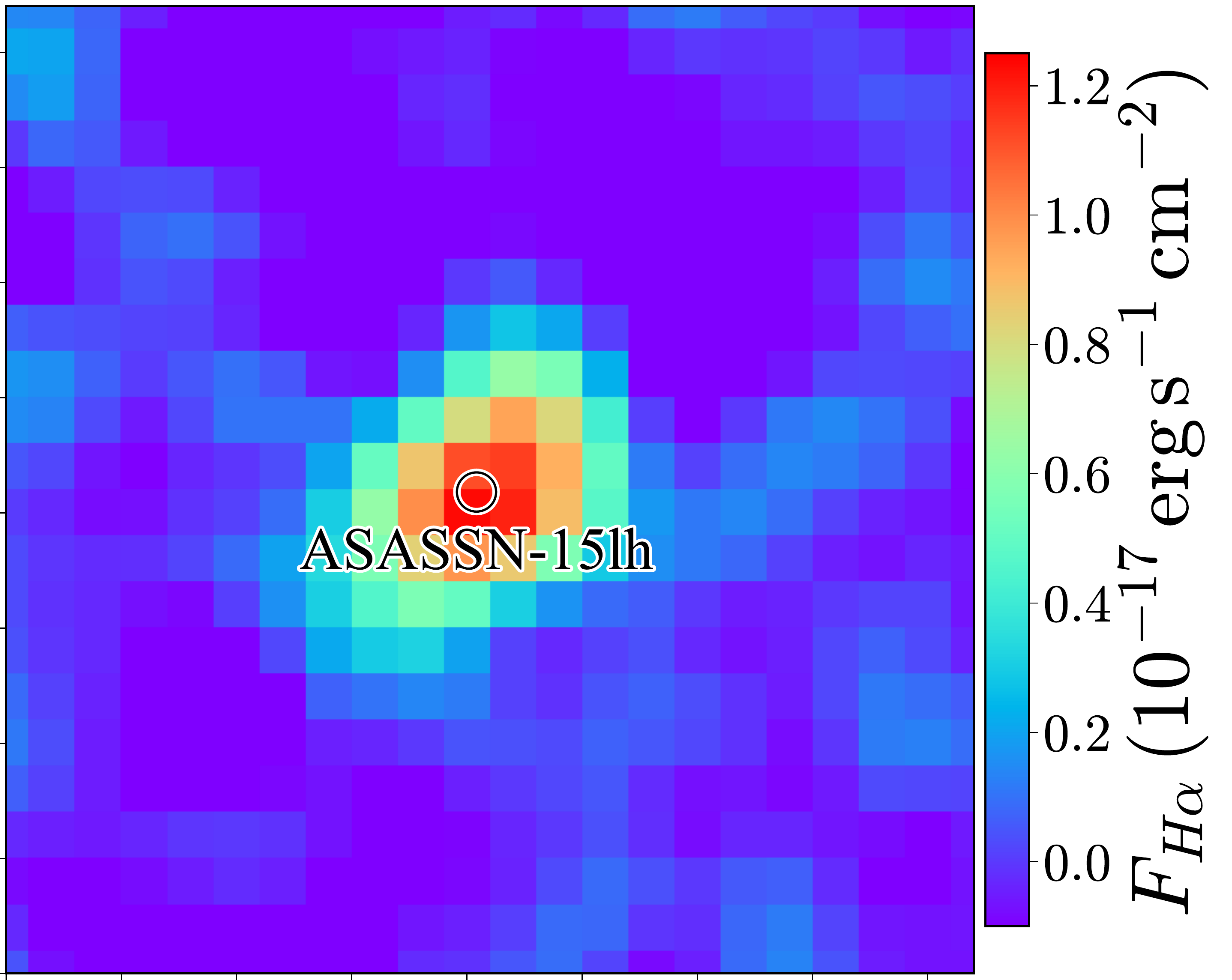}
\end{subfigure}
\begin{subfigure}{.2425\textwidth}
  \includegraphics[width=1.0\linewidth]{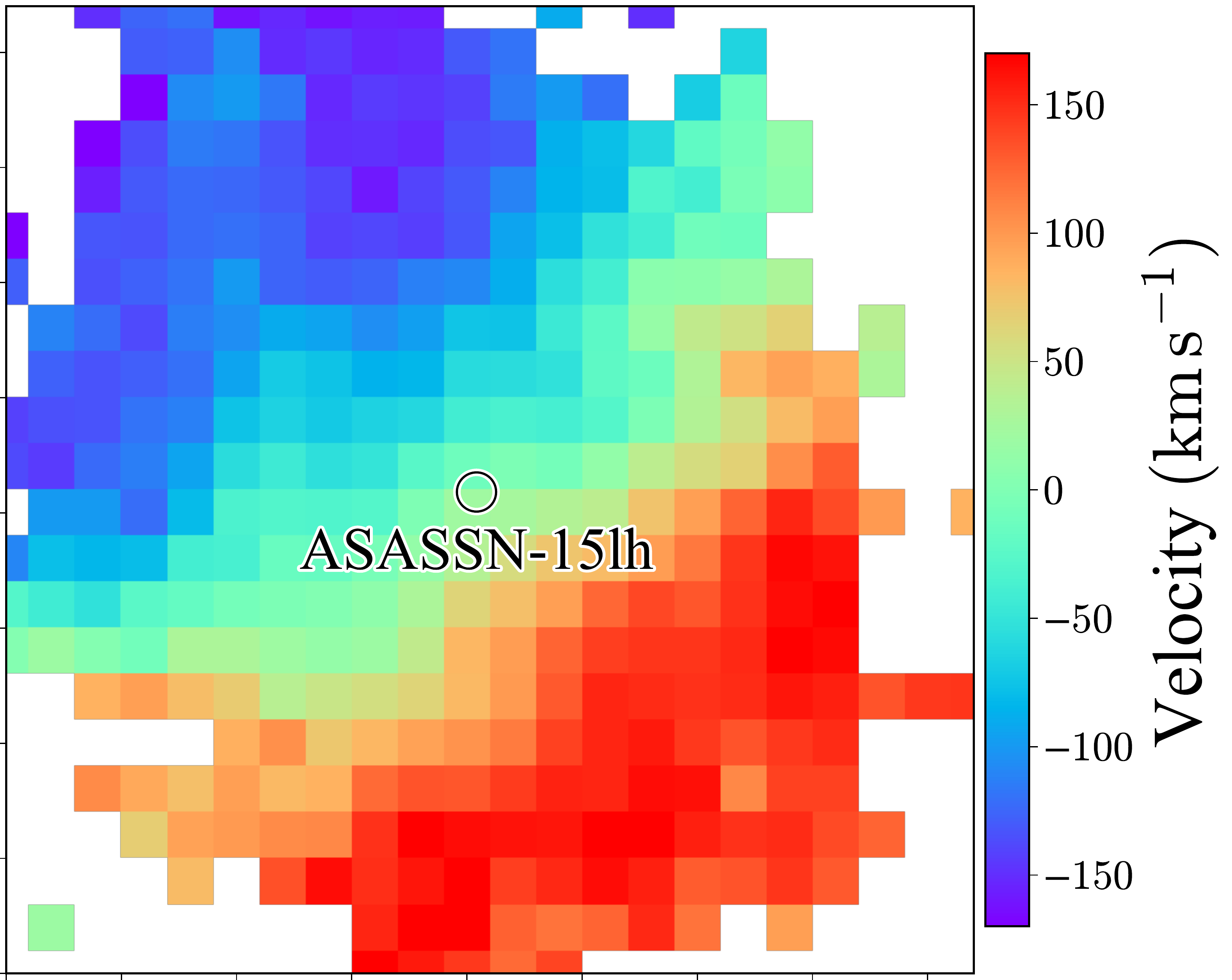}
\end{subfigure}
\begin{subfigure}{.2425\textwidth}
  \includegraphics[width=1.0\linewidth]{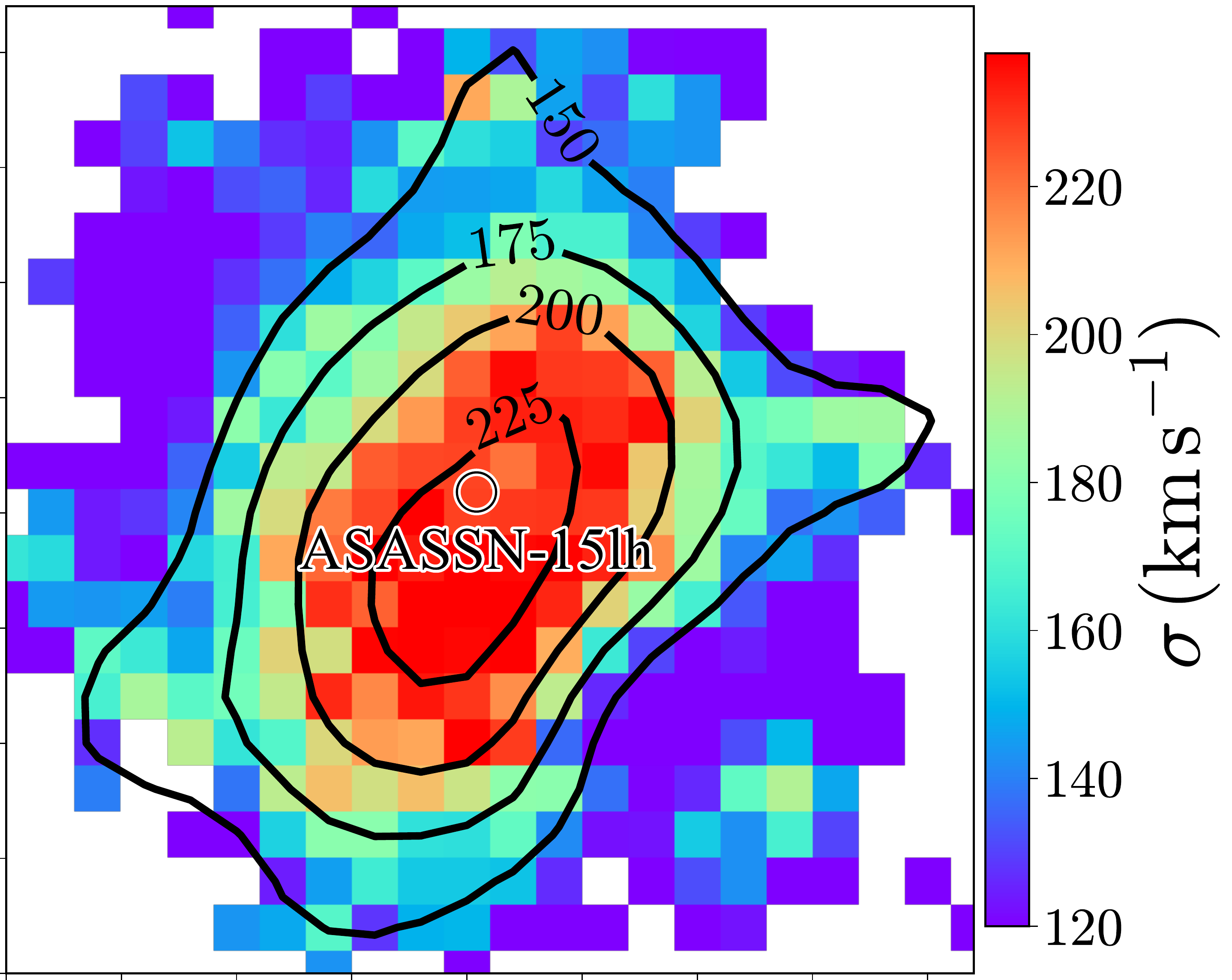}\end{subfigure}
\caption{{Spatially resolved properties of the ASASSN-15lh host. In the top row, reconstructed images of total intensity in the MUSE spectral range ($4650$~\AA~ to $9300$~\AA) in the \textit{left panel}, and the central component of \ha~emission in the \textit{right panel}. The bottom row shows the stellar velocity field \textit{(left panel)} and velocity dispersion $\sigma$ \textit{(right panel)}. Each of these images is 6\farc{2} by 6\farc{2}, which corresponds to 23~kpc by 23~kpc at the redshift of ASASSN-15lh. North is up and east is to the left}.}
\label{fig:sig_maps}
\end{figure}

A similar procedure as in the previous section (Section~\ref{sec:pPXF}, and Fig.~\ref{fig:stargas}) also returns the stellar kinematics by broadening and shifting template spectra until they match the observed data. We are in particular interested in the observed broadening of absorption lines to derive the velocity dispersion of the stellar component within the effective radius ($\sigma_{\mathrm{e}}$) because it correlates well with the mass of the central black hole \citep{2000ApJ...539L...9F, 2000ApJ...539L..13G}. Figure~\ref{fig:stargas_sig} shows medium-resolution X-Shooter and MUSE spectra (instrumental resolution in this wavelength range $\sigma_{\mathrm{inst}}\sim25$\,\kms~for X-shooter and $\sigma_{\mathrm{inst}}\sim80$\,\kms~for MUSE) in the wavelength range of the strong \ion{Ca}{II}~H+K doublet, \hd, \hg, and the G band, as well as the best fit continuum. The shown fits (Fig.~\ref{fig:stargas_sig}) result in a luminosity-weighted line-of-sight velocity dispersion of $\sigma=225\pm15$\,\kms, and fits to the data excluding the Balmer lines yield comparable values within errors. {The effective radius $R_{\mathrm{e}}$ of the galaxy is 0\farc{4}, and the measured luminosity-weighted velocity dispersion within the X-shooter slit of 1\farc{0} or $2.5 \times R_{\mathrm{e}}$ can thus be considered as a fair proxy of $\sigma_{\mathrm{e}}$ \citep{2000ApJ...539L..13G}. Measurements within $R_{\mathrm{e}}$ from the MUSE data cube (Figs.~\ref{fig:stargas_sig}~and~\ref{fig:sig_maps}) also show good agreement to the X-shooter value.}

This value of $\sigma_{\mathrm{e}}$ corresponds to a mass $M_\bullet$ of the central black hole of $M_\bullet = 5.3_{-3.0}^{+8.0}\cdot10^{8} M_\sun$ \citep[Eq. 3, 5, or 7 in][]{2013ARA&A..51..511K}. The quoted uncertainty is dominated by the intrinsic scatter in the $M_\bullet$-$\sigma$ relation ($\sim0.3$\,dex), and errors in measuring and inferring $\sigma_\mathrm{e}$ from our data do not contribute significantly. This mass estimate is similar to those obtained previously \citep{2015ATel.7776....1P, 2016NatAs...1E...2L}, but has somewhat smaller uncertainties. The Eddington luminosity for this black hole mass is $L_{\mathrm{Edd}}=7_{-4}^{+10}\cdot10^{46}$~erg~s$^{-1}$, which means that ASASSN-15lh radiated at $\sim 5$~\% of the Eddington ratio of the SMBH ($L_{\mathrm{peak}}\sim 3 \cdot 10^{45}$~erg~s$^{-1}$) at peak. This value is within the distribution of Eddingtion ratios of previously studied TDEs  \citep{2017ApJ...842...29H, 2017arXiv170608965W}.

{Figure~\ref{fig:sig_maps} summarizes galaxy properties through four maps of total intensity in the MUSE wavelength range (observed $4650$~\AA~ to $9300$~\AA), flux in the central component of the \ha~emission (Sections~\ref{sec:pPXF} and \ref{sec:prof}), line-of-sight velocity field $V$, and velocity dispersion\footnote{The galaxy kinematics were derived after summing nine adjacent spaxels to obtain a sufficiently high S/N, and pixels in the $\sigma$ and $V$ maps are hence correlated.}. The peak of the total intensity and the \ha~flux aligns well with the ASASSN-15lh transient, and the velocity field shows clear rotation in the direction of the photometric major axis (NE-SW direction). The stellar velocity dispersion rises toward the center and is somewhat elongated along the galaxy minor axis. Its center is slightly ($\sim$0\farc{2}$\pm$ 0\farc{1}) offset\footnote{The total error is a combination from uncertainties in the measurement of $\sigma$, errors during centroiding, and the correlation between individual spaxels due to seeing and spectral extraction} from the peak of the total intensity, which can be interpreted as a sign that the galaxy center has been disturbed by a past merger, but the significance of the offset is low. In general, the central velocity dispersion field of early-type galaxies is very rich in features \citep{2004MNRAS.352..721E} and our limited spatial resolution (FWHM=3~kpc) prevents us from making stronger claims with respect to the origin, nature, and implications of the velocity dispersion field.}

\subsection{Central stellar population}
\label{sec:spop}

\begin{figure}
  \includegraphics[width=0.999\linewidth]{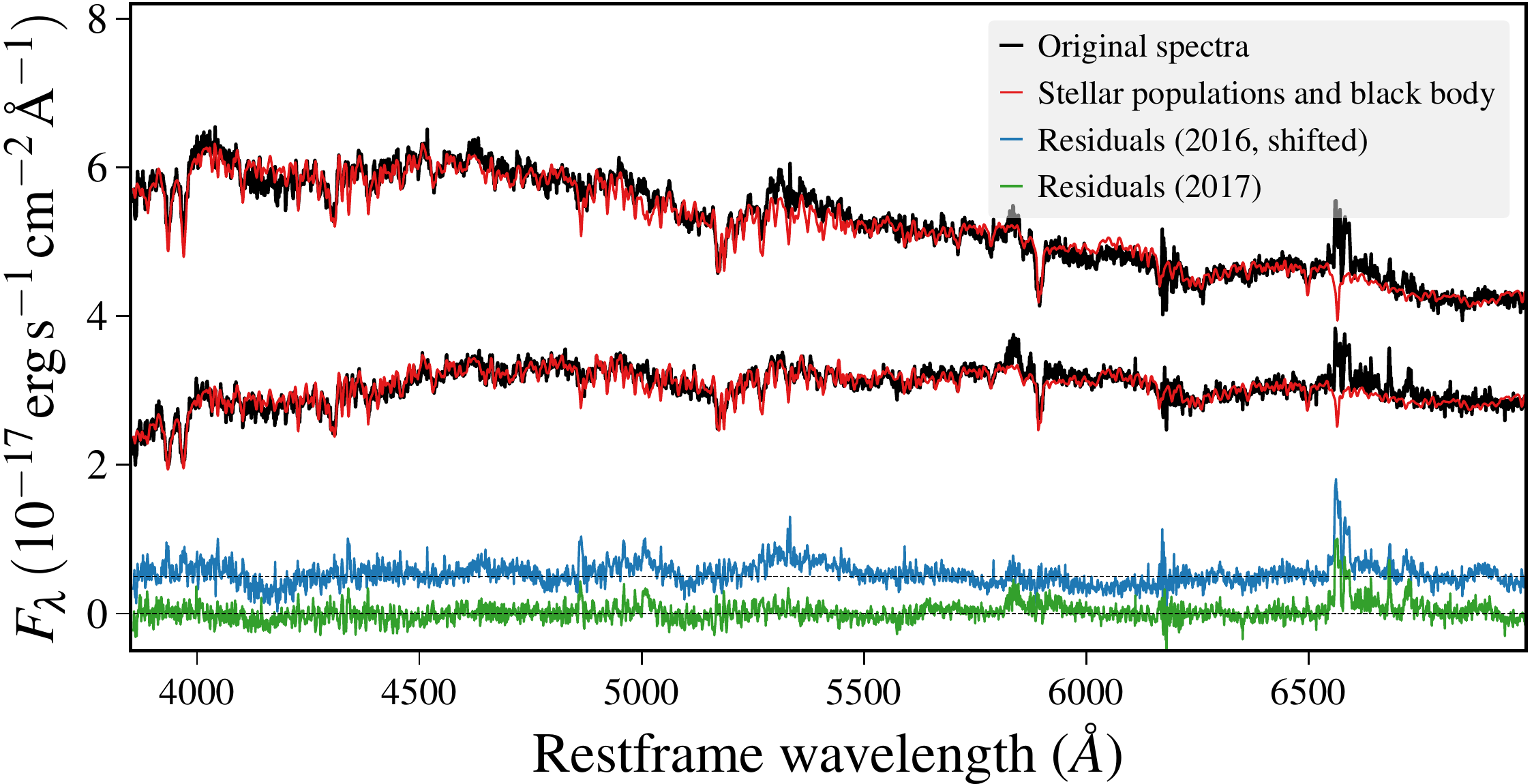}
\caption{{Physical model of transient plus multiple stellar populations fitted to the nuclear spectra (radius of 0\farc{5}) in the MUSE wavelength range. The 2016 data returns broad features in the residuals, for example at 4200~\AA~or 5300~\AA, which are less pronounced in the 2017 data. These residuals represent hence likely deviations from the transient spectrum from a pure blackbody.}}
\label{fig:stell_pop}
\end{figure}

{We also performed a second set of fits to derive a more physical interpretation of the continuum by modeling the transient with a blackbody component (as opposed to the low-order polynomial in Section~\ref{sec:pPXF}) and the galaxy with a superposition of templates from single stellar populations (Fig.~\ref{fig:stell_pop}). These fits with a blackbody continuum turned out to be less accurate in disentangling emission lines from the continuum as they returned broad residuals in the subtracted spectrum}. The residuals are likely deviations of the ASASSN-15lh spectrum from a pure blackbody, as similar fits to spectra of the EI and EII galaxies (Fig.~\ref{fig:fc}) show no residuals, and they are much weaker in the 2017 epoch (Fig.~\ref{fig:stell_pop}). Here, we used \texttt{starlight} \citep{2005MNRAS.358..363C, 2009RMxAC..35..127C} and \citet{2003MNRAS.344.1000B} templates in a similar manner as we described in detail elsewhere \citep{2016MNRAS.455.4087G, 2017arXiv170205430K}.

The detailed physical parameters of the central stellar population vary somewhat depending on the exact choice of the size of the spectral extraction, the best fit transient blackbody temperature, and the base list of galaxy templates. However, contributions from two stellar populations are always required for a satisfactory fit of the stellar component in the galaxy center: a dominant population with an age of around 1~Gyr to 2~Gyr, which typically contributes roughly 60\% to 80\% to the composite galaxy spectrum, and a second population that is significantly older with an age between 10~Gyr and 13~Gyr and contributes between 10\% and 30\% of the stellar light\footnote{Previously estimated stellar population ages of around 5~Gyr for the ASASSN-15lh host were based on pre-explosion broadband photometry \citep{2015ATel.7843....1M, 2016NatAs...1E...2L,  2016Sci...351..257D}, and as such only probe the integrated light of many stellar populations.}. The presence of a 1-2 Gyr old population is compatible with the oldest TDE hosts studied previously \citep{2016ApJ...818L..21F, 2017ApJ...835..176F}. 

The strength of the \hd~absorption line has often been used for a post-starburst, or E+A, classification in galaxies \citep{1999ApJS..122...51D, 1997ApJS..111..377W}, and in particular galaxies with TDEs \citep[e.g.,][]{2014ApJ...793...38A, 2016ApJ...818L..21F, 2017arXiv170701559L}. For the ASASSN-15lh host, we estimate an EW$_{\mathrm{H}\delta}\sim 2~\AA$~by measuring the optical depth of the \hd~absorption in the spectra and comparing it to the continuum from galaxy models derived via pre-explosion photometry\footnote{Even our latest spectra from 2017 still show a weak contamination from the transient so that a direct application of the Lick indices is not possible.} at the respective wavelength. Following \citet{1992ApJS...78....1D}, an EW$_{\mathrm{H}\delta}\sim2~\AA$~together with the absence of \oii($\lambda 3727$)~in our spectra (Figure~\ref{fig:stargas_sig}) classifies the host as a passively evolving galaxy.

\subsection{Emission-line profiles}
\label{sec:prof}

Once the continuum from stars and transient has been separated from the emission lines, we look more closely to the line shape and velocity structure of the \ha\, and \nii\, complex of Fig.~\ref{fig:stargas} to study the ionization source and its relation to the transient ASASSN-15lh itself.

\begin{figure}
  \includegraphics[width=0.999\linewidth]{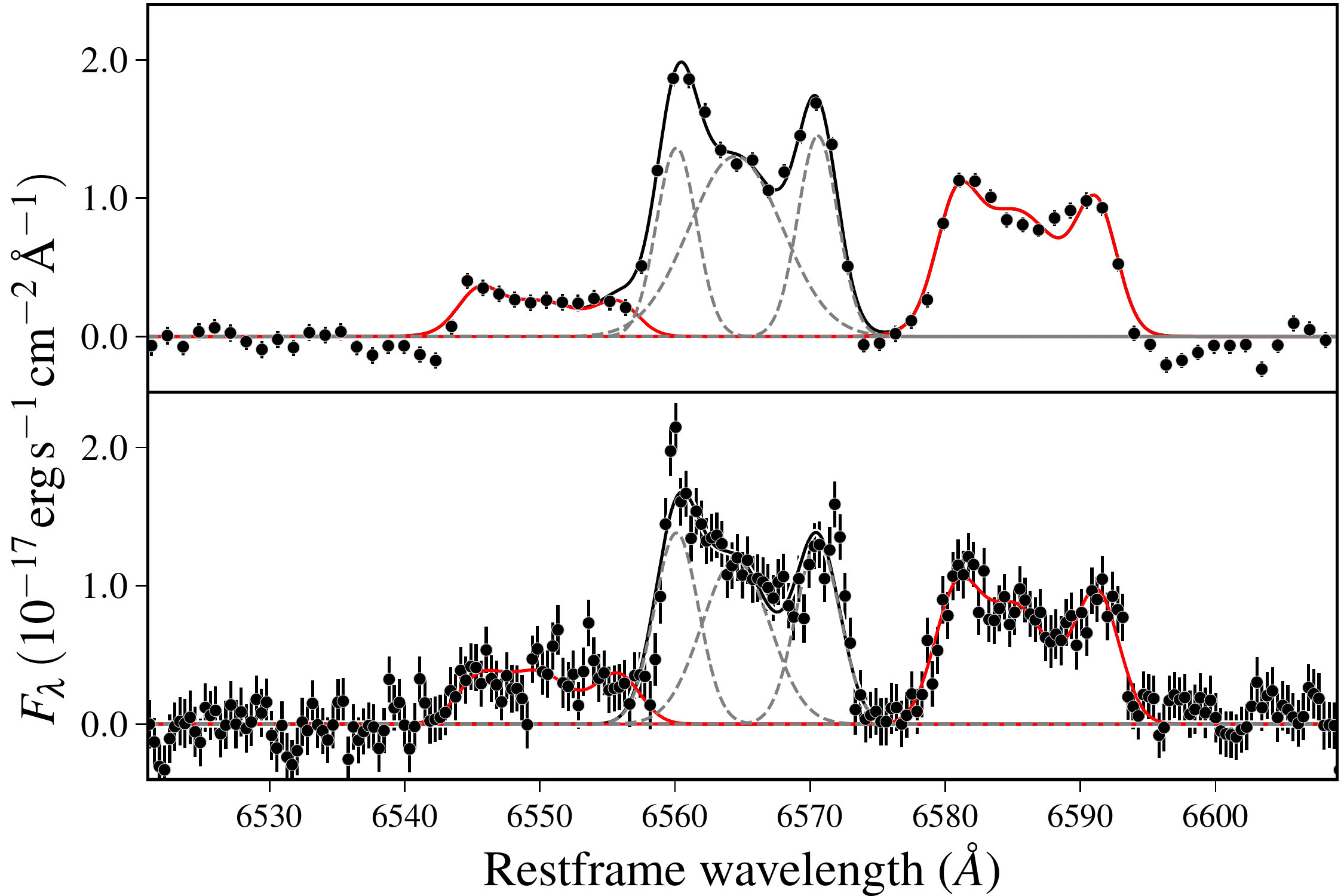}
\caption{Emission-line profiles and component decomposition for the \ha\, and \nii($\lambda$6548, 6584)\, complex. The \textit{upper panel} shows a spectrum extracted for the central component (radius of 0\farcs{9} or 3.4~kpc around the galaxy center) of the host galaxy from the MUSE data cube; the \textit{lower panel} shows the X-Shooter spectrum. The contribution from \nii~ is shown in red; the total sum of the emission-line model in black. For \ha, we also plot the three individual Gaussian components with dashed lines.}
\label{fig:hanii}
\end{figure}

In Fig.~\ref{fig:hanii}, we show a zoom in on the continuum-subtracted wavelength range of the respective transitions from our MUSE and X-Shooter spectroscopy. The line shape of the individual transitions is complex, and each line transition is composed of three components that are well separated in wavelength space: a central, broader component ($\mathrm{FWHM}=8$~\AA~ or 380~\kms) and narrower ($\mathrm{FWHM}=3$~\AA~ or 100~\kms) blue and red components offset by approximately 5~\AA~ (250~\kms) in each direction. The redshifts of the three components are $z=0.2310$, $z=0.2318,$ and $z=0.2331$ with errors of about $\pm$0.0002 each. 

The originally reported redshift was $z=0.2326$ from narrow \ion{Mg}{ii} absorption lines \citep{2015ATel.7774....1D}, which we also confirm with our spectra. However, \ion{Mg}{ii} absorption does not necessarily yield the most accurate galaxy redshift, as the \ion{Mg}{ii} gas clouds are subject to random motion within the gravitational potential of the galaxy. Hence, in the following, we adopt $0.2318\pm0.0002$ measured in a heliocentric reference frame for the systemic redshift of the host from galaxy emission and stellar absorption lines. This change is insignificant for this or any of the previous articles about ASASSN-15lh.

The line shape of the \ha\,line and each of the collisionally excited \nii($\lambda\lambda$6548, 6584) lines is identical within the measurement uncertainties. Also, the line shape is well resolved, in particular through the medium-resolution X-Shooter data and appears comparable between the two spectrographs\footnote{The width of the instrumental resolution at the wavelength range of \ha~(observed 8100\,\AA) has a FWHM$\sim110$~\kms~ for MUSE and a FWHM$\sim35$~\kms~for X-Shooter.}.

The emission lines from the \hb~ and \oiii~ transitions are generally consistent with this picture (Fig.~\ref{fig:stargas}), in particular \hb~ shows evidence for a similar line profile. However, large statistical errors stemming from the bright background of galaxy and transient and systematic uncertainties from the continuum subtraction prevent us from performing a detailed kinematic analysis for any other lines except \ha~and \nii.

To derive line fluxes and gas kinematics, we fit a superposition of three Gaussians for each of the three transitions simultaneously to the spectra shown in Fig.~\ref{fig:hanii}. The intrinsic line width, broadened by the instrumental resolution, is tied between both instruments and the three transitions, and only the normalization is allowed to vary during the fit; while the MUSE spectra should measure the full flux of the emission lines, the slit from X-shooter might lead to slit losses. 

\subsection{Positional analysis of the emission-line components}
\label{sec:posana}

The MUSE integral-field spectroscopy allows us to go beyond a standard kinematic analysis as carried out above and perform a spatially resolved analysis of the individual velocity components. Here, we created a continuum-subtracted data cube from the original MUSE spectroscopy by performing a fit similar to that of Fig.~\ref{fig:stargas}, but now for each individual spaxel in the astrometrically calibrated data (Sect.~\ref{obs:muse}). Similar procedures have been used by us frequently in the past on MUSE data \citep{2016MNRAS.455.4087G, 2016ApJ...830L..32P, 2017arXiv170205430K}, and allow us to combine and visualize the spatial information of the MUSE maps with the velocity information of the emission-line kinematics.

\begin{figure}
\begin{subfigure}{.49\textwidth}
  \includegraphics[width=0.999\linewidth]{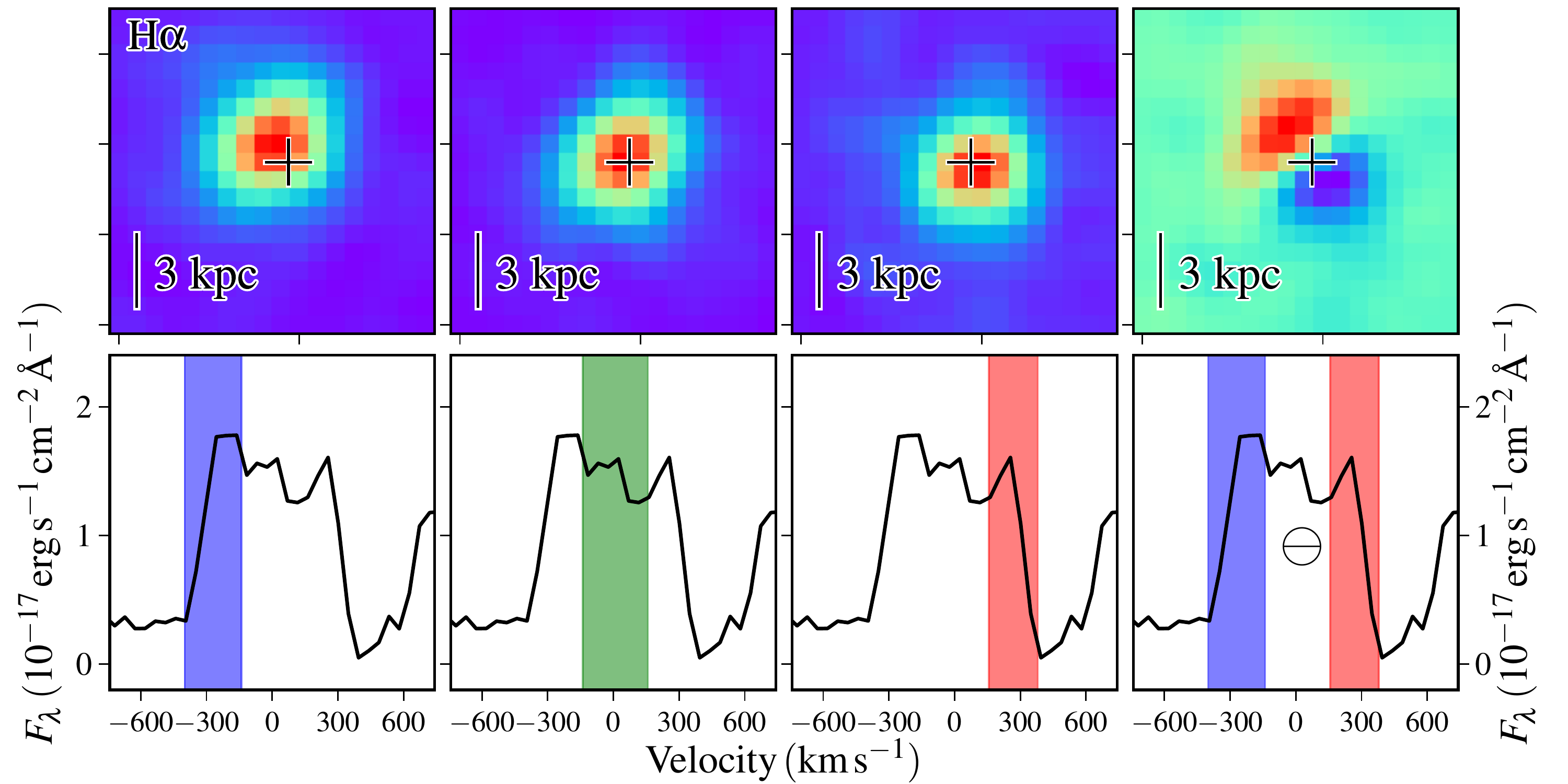}
\end{subfigure}
\begin{subfigure}{.49\textwidth}
  \includegraphics[width=0.999\linewidth]{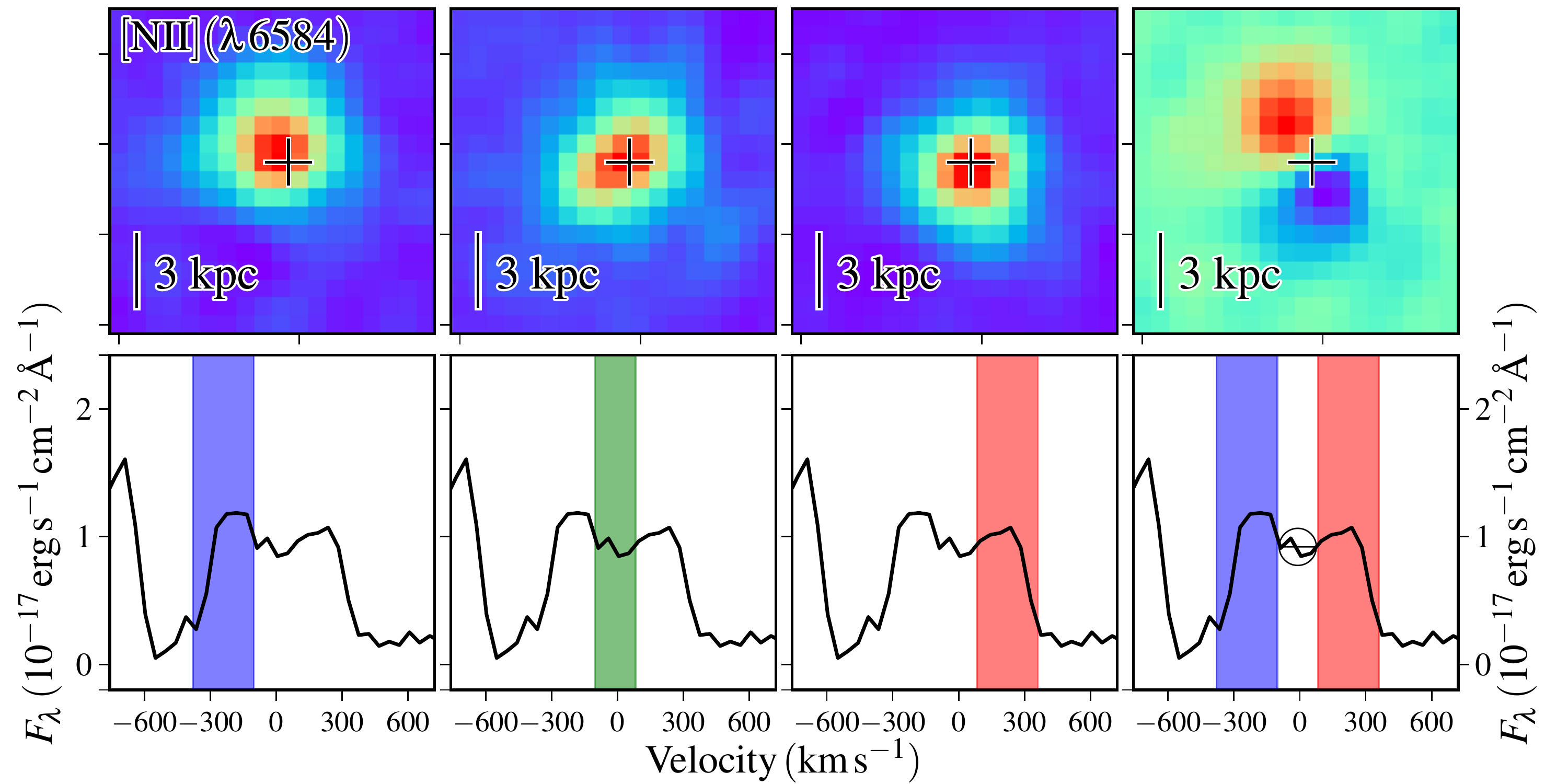}\end{subfigure}

\caption{Channel maps of the velocity components. The two \textit{upper} rows show in the three leftmost panels the reconstructed images in a given velocity range (indicated by the colored region in each spectrum) from the continuum-subtracted MUSE map for \ha. The right panel shows a subtraction between the blue and red components. The position of the transient is indicated by a cross. The two \textit{lower} rows show the same for \nii. All images have been smoothed by a Gaussian kernel with a size of the FWHM of 2~px by 2~px for illustration purposes. The physical scale of the images is indicated by the bar in the lower left corner of each image. North is up and east is to the left in all images.}
\label{fig:channelmaps}
\end{figure}

Figure~\ref{fig:channelmaps} shows the channel maps of the continuum-subtracted spectroscopy at the center of the ASASSN-15lh host and in the wavelength range of \ha. Each of the three panels shows the reconstructed image in the given wavelength range, whereas the rightmost panel is a subtraction between the bluest and reddest component. The position of the transient as derived through the HST-to-MUSE astrometric alignment is indicated by a cross.

It is evident that the three velocity components are not only separated in velocity space, but are also located at different positions. In addition, the blue and red components are offset from the transient location. The velocity separation between the blue and red component is 500~\kms, and the spatial offset is $1.8\pm0.3$ MUSE spaxel or $0\farc{36}\pm0\farc{06}$, corresponding to a projected distance of $1.3\pm0.2$~kpc. 

The central component is placed symmetrically between the red and blue emission peaks (both spatially and in velocity space), and is consistent with the transient position within the combined astrometric uncertainty of $0.22$ MUSE spaxels, or 45 mas, which corresponds to a physical scale of 170~pc at $z=0.232$.

\subsection{Ionization source}
\label{sec:ionsource}

The strength and ratios of various collisionally excited and recombination lines of metal and hydrogen ions trace the physical conditions in the gas phase and the origin of the radiation that was ionizing the gas in the first place. A useful and widely applied diagnostic plot is the Baldwin-Phillips-Terlevich (BPT) diagram \citep{1981PASP...93....5B}, which discriminates between ionizing flux coming from the hard radiation of AGN or shocks or \hii~regions where the UV flux is dominated by massive stars through the ratios of \nii/\ha~ and \oiii/\hb.

The BPT diagnostic is frequently used for host galaxies of transient objects and various classes of objects occupy very different phase spaces. For example, the hosts of $\gamma$-ray bursts or superluminous supernovae typically reside in the high \oiii/\hb, low \nii/\ha~regime \citep{2015A&A...581A.125K, 2015MNRAS.449..917L}, characteristic of young starbursts at low metallicity. In contrast, the nearby ($d_L=90$~Mpc) TDE ASASSN-14li \citep{2016MNRAS.455.2918H} has shown an extended structure of ionized gas with emission-line ratios that imply ionization from an AGN \citep{2016ApJ...830L..32P}.

\begin{figure}
  \includegraphics[width=0.999\linewidth]{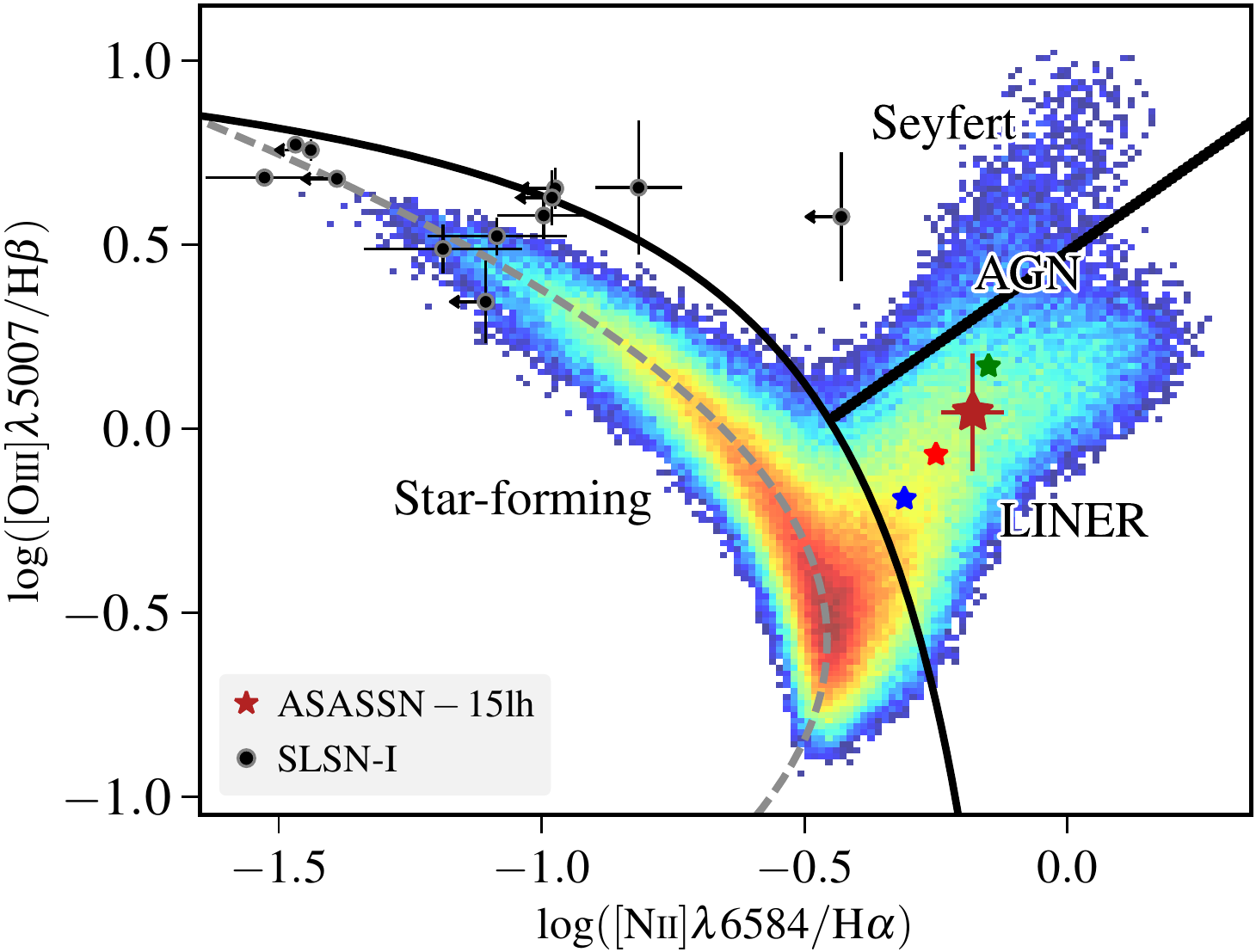}
\caption{Host of ASASSN-15lh in the BPT diagram. The large red star represents the total integrated flux, while the three smaller stars show the three individual components in their respective color, i.e., the green is the central component. Error bars of the individual components are not shown to enhance clarity in the figure. To indicate the size of the respective error bars, the location of the smaller stars is consistent with each other within 1 $\sigma$. The black data points indicate hydrogen-poor SLSNe from \citet{2015MNRAS.449..917L} with limits indicated by arrows. The black solid lines represent differentiation lines between star-forming galaxies and AGN from \citet{2013ApJ...774L..10K} at $z\sim0.23$, and between LINERs and Seyferts from \citet{2010MNRAS.403.1036C}. The gray dashed line indicates the ridge line, i.e., the line with the highest density of star-forming galaxies in SDSS \citep{2008MNRAS.385..769B}.}
\label{fig:BPT}
\end{figure}

The ionized regions of ASASSN-15lh are plotted in the BPT diagram in Fig.~\ref{fig:BPT}, where we use the SDSS DR7 spectroscopy \citep{2009ApJS..182..543A} with line fluxes from the MPA/JHU catalog as a background sample. The total fluxes of the four emission lines (\ha, \hb, \nii($\lambda$6584), and \oiii($\lambda$5007)) and thus their ratios are rather well constrained, in particular when adding the constraint that \oiii($\lambda$5007)/\oiii($\lambda$4959)=3 in the fit. However, for the \oiii~and~\hb~lines, the individual components are not easy to separate and the flux ratio has hence large uncertainties.

It is evident that all individual components of the line emission, and their integrated flux, are located in the part of the BPT diagram that is occupied by low-luminosity AGN and shock ionization or excitation, similar to many other TDE hosts \citep{2017ApJ...835..176F}. And even though the measurement error, especially for the \oiii/\hb\ ratio is substantial, it is clear that all components occupy a region in the plot that is offset from the star-forming sequence of SDSS galaxies.

In particular the central component, which is positionally coincident with the transient, has a high value of \nii/\ha, inconsistent with pure star formation. Our measurements of line ratios hence require that at least a significant fraction of the ionization is coming from AGN or shocks \citep{2011MNRAS.413.1687C}. Similarly, a classification based on the equivalent width (EW) of \ha  \ and the \nii/\ha~ratio \citep{2011MNRAS.413.1687C} shows the central region of the ASASSN-15lh host to be consistent with a weak AGN (\nii/\ha~$>0.4$, and $3~\AA < \mathrm{EW}_{\mathrm{H}\alpha} < 6~\AA$), sometimes referred to as low-ionization nuclear emitting regions or LINERs. 

Given that the \ha~emission is more compact than the stellar emission, the exact value of $\mathrm{EW}_{\mathrm{H}\alpha}$ depends on the size of the spectral extraction. Using central spectra extracted in the region of the line emission (0\farc{9} radius), $\mathrm{EW}_{\mathrm{H}\alpha}=3.2\pm0.4$~\AA, while it is $\mathrm{EW}_{\mathrm{H}\alpha}=1.3\pm0.2$~\AA~when considering the full extend of the galaxy.

The leading theoretical model to explain LINERs is photoionization from a central, low-luminosity AGN \citep[e.g.,][for a review]{2008ARA&A..46..475H}. In contrast to other ionization sources (young stars, fast shocks, and evolved stars), a narrow-line region (NLR)\ of a central AGN would naturally explain the line ratios, the kinematics, and, as shown in Section~\ref{sec:nation}, the spatial offset between the three observed kinematic components seen in the host of ASASSN-15lh.

We conclude that the observed line emission is most likely the radiation from gas photoionized by AGN radiation and neither from star formation nor the transient itself. Deriving exact limits on the SFR from the \ha~emission is complex. While we can exclude from the BPT diagram that all of the ionized gas emission comes from star formation, we cannot strictly exclude this for a smaller fraction. Assuming that the star formation powered \ha~line flux is less than half of the observed flux, the star formation rate (SFR) would be $SFR \lesssim 0.1$~\Msunyr \citep{1998ARA&A..36..189K}. This implies that the host is quiescent in terms of star-formation, lying at least two orders of magnitude below the SFRs of similarly massive galaxies on the main sequence \citep[e.g.,][]{2010ApJ...721..193P, 2012ApJ...754L..29W}.

\subsection{Constraints on the bolometric AGN luminosity}

Summing over all spectral components, the total \oiii~or \ha~luminosity of the gas photoionized by the AGN is $L_{\oiii}\sim10^{40}$~erg~s$^{-1}$ or $L_{\mathrm{H}\alpha} \sim 3 \times 10^{40}$~erg~s$^{-1}$, respectively. These values correspond to an X-ray luminosity of around $L_{\mathrm{X}}\sim10^{41}$~erg~s$^{-1}$ in the 2-10 keV energy range, or a bolometric luminosity of $L_{\mathrm{bol}}\sim10^{42} - 10^{43} $~erg~s$^{-1}$ assuming average correction factors \citep{2008ARA&A..46..475H, 2009A&A...504...73L, 2012MNRAS.425..623L}. Even though these luminosity estimates are uncertain by at least an order of magnitude, the Eddington ratio $\lambda_{\mathrm{edd}}$ of the central black hole is far below unity ($\lambda_{\mathrm{edd}}\sim10^{-5}-10^{-4}$) for the AGN emission.

The inferred X-ray luminosity of a weak AGN is comparable to the emission observed by \citet{2017ApJ...836...25M}, who estimate a 0.3-10~keV luminosity of $L_{\mathrm{X}}\sim10^{41} - 10^{42} $~erg~s$^{-1}$. Given that this measurement is derived in a somewhat larger energy interval, and the X-ray source seems relatively soft, we cannot rule out that the observed X-ray flux is caused by steady accretion of a low-luminosity, pre-existing AGN. Given that the pre-transient, AGN-dominated X-ray emission estimated here, and the recent X-ray detection \citep{2017ApJ...836...25M} are both uncertain by at least an order of magnitude, the available constraints on X-ray fluxes are also fully consistent with substantial X-ray variability.

\section{Discussion}
\label{sec:Disc}

\subsection{Nature of the ionized-gas emission}
\label{sec:nation}

As described in the previous paragraphs (Section~\ref{sec:prof} and \ref{sec:posana}), the emission-line profile observed in the ASASSN-15lh host galaxy nucleus is complex and consists of multiple components with spatial and velocity offsets (Figures~\ref{fig:hanii} and \ref{fig:channelmaps}). The components are aligned along a single direction, and all are likely to be ionized by an AGN. The double-peaked nature of the \ha~ and \nii~line initially led us to suspect a binary AGN as origin of the emission. Giant early-type galaxies such as the ASASSN-15lh host are thought to be the result of galaxy mergers \citep[e.g.,][and references therein]{2006ApJS..163....1H}. Because supermassive black holes (SMBHs) arguably reside in the centers of all massive galaxies \citep[e.g.,][for a review]{2013ARA&A..51..511K}, a good fraction of spheroidal galaxies should also host binary SMBHs. If active, for example through nuclear accretion induced by the merger, the double SMBH appears as a binary AGN. The most convincing cases of binary AGN on kpc scales, as would be the case here, have been imaged as double point sources in hard X-ray \citep[e.g.,][]{2003ApJ...582L..15K, 2008MNRAS.386..105B} or radio emission \citep{2011ApJ...740L..44F, 2015ApJ...813..103M}.

However, this scenario seems unable to fully explain our observations. While the line shape from our MUSE data could conceivably be explained with two components only, the higher resolution X-Shooter data clearly demonstrates the presence of an even more complex structure (Fig. ~\ref{fig:hanii}). We would hence need to invoke three aligned and active SMBH, which we consider too contrived to explore any further. Only the medium-resolution X-Shooter data (FWHM=35\,\kms) has allowed us to convincingly rule out this possibility. Based on the spectral resolution of  MUSE (FWHM=150\,\kms) alone, we would have probably considered a binary AGN as the cause of the line profile more seriously.

Instead, narrow-line gas kinematics in a rotating disk or galactic winds driven by the AGN offer much more natural explanations for the observed kpc-scale emission in Balmer and collisionally excited metal lines \citep[e.g.,][]{2011ApJ...735...48S}. Indeed, complex narrow line regions (NLRs) are explained with AGN outflows at least for some nearby galaxies \citep[e.g.,][]{2011ApJ...727...71F} but the differentiation between rotating NLR disks and genuine outflows is not always trivial \citep{2011ApJ...735...48S, 2015ApJ...813..103M} in general. In our case, we observe a relatively symmetrical line profile and a geometry that is aligned along the major axis of the galaxy (Figs.~\ref{fig:fczoom} and~ \ref{fig:channelmaps}), both of which argues for an origin in a rotating disk of ionized gas. Clearly, the stellar population is much more extended than the emission lines (Figure~\ref{fig:sig_maps}), and neither the red nor blue component of the emission lines has a kinematic counterpart to absorption lines in the core. However, the stellar velocity field corotates with the ionized gas on larger scales (Figure~\ref{fig:sig_maps}). We hence suggest that the emission-line shape and position of all components is caused by an AGN at the position of the central component with a rotating disk of ionized gas explaining the velocity and spatial structure. {The observed velocity and radius of the narrow-line disk then constrains the enclosed mass $M_\mathrm{enc}$. Assuming that the disk is viewed nearly edge-on, using a circular rotation velocity of $v_{\mathrm{H}\alpha}\sim250$~\kms, radius $r_{\mathrm{H}\alpha}\sim650$~pc (Section~\ref{sec:posana}) and following \citet{2003ApJ...592...42G}, we derive $M_\mathrm{enc}\sim 9 \times 10^{9}$~\Msun, a factor 20 larger than the mass of the SMBH, and a factor 10 lower than the total stellar mass of the galaxy \citep{2016NatAs...1E...2L}.}

{The \oiii~luminosity of Seyferts correlates with the size of the NLR \citep[e.g.,][]{2003ApJ...597..768S}. This correlation would predict a radius of 200~pc, significantly smaller than the observed radius of 650~pc. Or conversely, the observed NLR radius corresponds to a \oiii~luminosity of $L_{\oiii}\sim10^{41.5}$~erg~s$^{-1}$. This mismatch possibly indicates that the central ionizing source was more active in the past, leaving behind only an extended NLR.}

\subsection{Nature of the transient}

Having pinpointed the location of the transient to the position of a weak AGN (and therefore a supermassive black hole), it is reasonable to relate both of these phenomena. The ASASSN-15lh environment thus constrains the nature of ASASSN-15lh itself with three different scenarios typically discussed in the pertinent literature in similar cases \citep[e.g.,][]{2011ApJ...741...73V, 2011ApJ...735..106D, 2014MNRAS.445.3263H, 2015ApJ...798...12V}: a very luminous core collapse supernova in the nuclear region of the host, a tidal disruption event, or intrinsic variability from the AGN.

The first of those, a luminous supernova at the nucleus of a passive galaxy, is somewhat contrived for ASASSN-15lh in several ways: firstly, there are the inconsistencies of the spectral properties and temporal evolution of the transient with other SLSNe or luminous SNe of type IIn as mentioned in the introduction or discussed in \citet{2016NatAs...1E...2L}. In addition, the association of a massive-star related phenomenon with an early-type host, a $\sim$Gyr old stellar population, and no obvious signs of recent star formation does not seem very viable. Finally, the positional coincidence with a NLR of a weak AGN and its location in the BPT diagram (Fig.~\ref{fig:BPT}) strongly suggest that ASASSN-15lh is the result of a physical phenomenon closely related to the SMBH and not star formation. 
AGN, and to some extent also non-active SMBHs such as that in the center of the Milky Way, are known to be variable throughout the electromagnetic spectrum on various timescales \citep[e.g.,][]{1997ARA&A..35..445U, 2001Natur.413...45B}. Typically, these forms of AGN variability are considered to be stochastic resulting from changes in the accretion rate, and are therefore clearly not able to explain the dramatic ASASSN-15lh variability and spectral evolution. However, much more extreme variations have been observed in changing-look quasars on shorter timescales \citep[e.g.,][]{2014ApJ...788...48S, 2015ApJ...800..144L, 2017ApJ...835..144G}, even though, in these cases as well, the discrimination between AGN activity and TDEs is not always trivial \citep{2015MNRAS.452...69M}. Two lines of evidence indicate that such an AGN-related event is not the origin of ASASSN-15lh itself. Firstly, no strong X-ray variability contemporaneous with large UV variability for ASASSN-15lh is observed, despite regular and simultaneous monitoring \citep{2016ApJ...828....3B, 2017ApJ...836...25M}. And secondly, the nebular line emission is constant in the various epochs of our spectroscopic monitoring. There is no evidence for appearing or disappearing of broad emission lines nor an increase in the continuum from a Seyfert 1 in our spectra, arguing against the interpretation of ASASSN-15lh as a changing-look quasar.

In the light of our detailed observation of the environment of the transient, it thus appears that the association of ASASSN-15lh with a TDE \citep{2016NatAs...1E...2L, 2017ApJ...836...25M} represents the most plausible explanation. Some TDEs within galaxies with regions of AGN-related ionization and excitation have been discovered before, where ASASSN-14li is the best-observed event \citep{2016Sci...351...62V, 2016MNRAS.455.2918H, 2016ApJ...830L..32P}. A more recent candidate is reported in \citet{2017arXiv170307816B}. Of course, a TDE is not the only explanation that would physically relate the transient with a SMBH in the center of a galaxy. \citet{2017arXiv170606855M}, for example, have proposed that some luminous nuclear transients within AGN are due to an interaction between accretion disk winds and clouds in the broad-line region. Because AGN broad lines are not detected in the ASASSN-15lh host, a direct application of this scenario to our situation does not seem straight forward, however, it serves to illustrate that other physical mechanisms apart from TDEs are still clearly interesting to explore for the remarkable transient emission observed as ASASSN-15lh.

\subsection{Spin of the supermassive black hole}

\begin{figure}
  \includegraphics[width=0.999\linewidth]{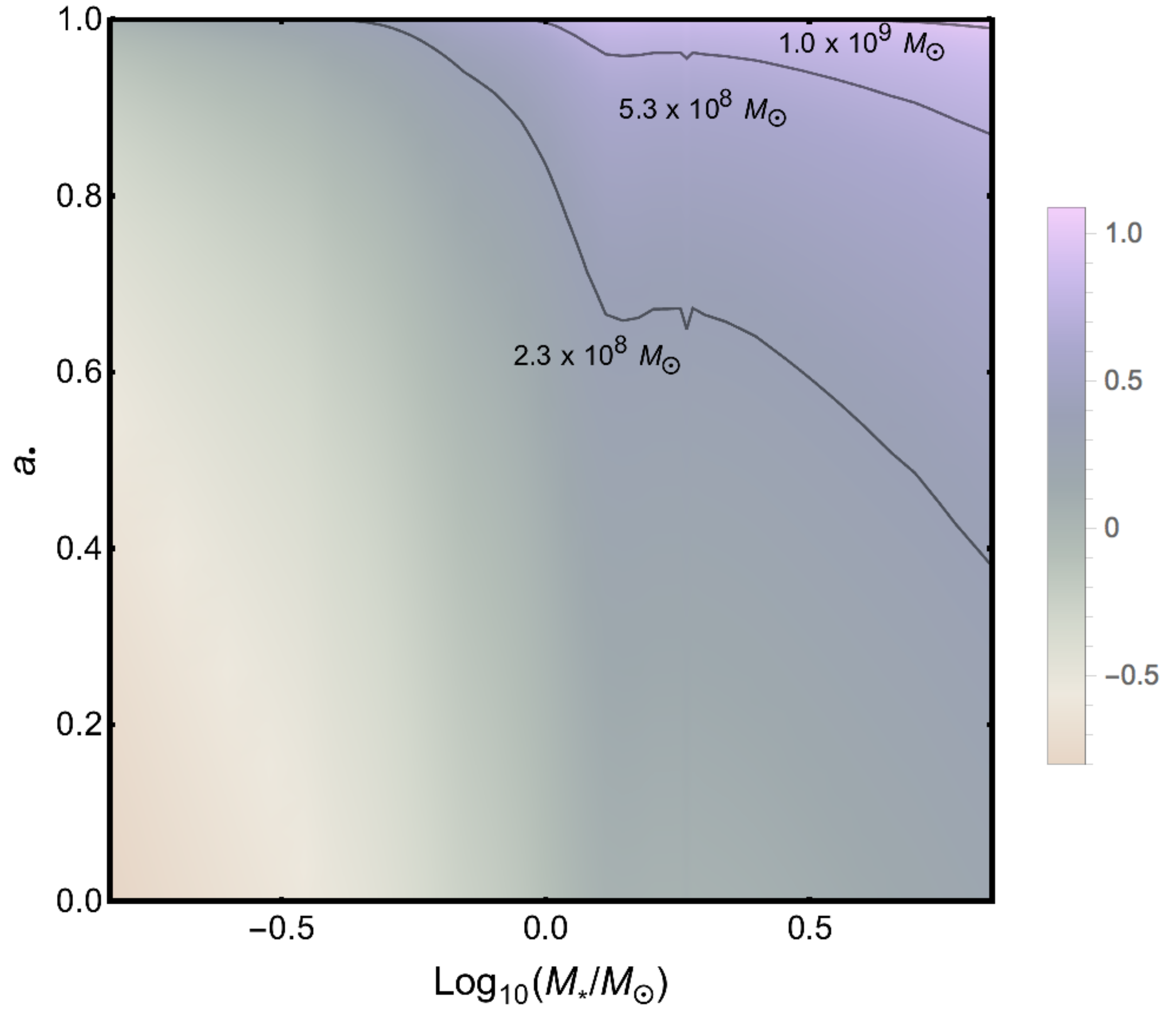}
\caption{Dependence between minimum spin parameter $a_\bullet$ (y-axis), mass $M_\star$ of the disrupted star (x-axis), and SMBH mass in $10^8~M_\odot$ in the color coding. Shown in black are also three tracks of specific SMBH masses, demonstrating that there is no viable solution for TDEs for zero black hole spin, or SMBH masses greater than $10^9~M_\odot$ within the constraints on the other parameters from this work. This plot assumes optimal orbital orientation between disruption axis and the spin vector, uses the ZAMS mass-radius relation for stars with $Z=0.008$ from \citet{2000A&AS..141..371G}, the formalism of \citet{2012PhRvD..85b4037K}, and follows the calculations detailed in the supporting material of \citet{2016NatAs...1E...2L}.}
\label{fig:Hills}
\end{figure}

As noted by other authors already \citep[e.g.,][]{2015ATel.7776....1P, 2016Sci...351..257D, 2016NatAs...1E...2L}, and shown in Figure~\ref{fig:Hills}, the SMBH mass of $M_\bullet = 5.3_{-3.0}^{+8.0}\cdot10^{8} M_\sun$ remains too high for a non-spinning black hole to disrupt stars with $M_\star<2.5$~\Msun~; i.e., the most massive stars with lifetimes shorter than the approximate age of the starburst from Section~\ref{sec:spop}. These parameters correspond to a limit\footnote{Based on the observational quantities and not taking into account that disruptions of lower mass stars are more likely, which would further increase the required Kerr parameter.} on the black hole spin $a_\bullet \gtrsim 0.5$ (Figure~\ref{fig:Hills}). {The cosmic evolution of the spin of SMBHs is driven by mergers and accretion, such that the required moderate to high spin parameter might not be uncommon in post-merger galaxies \citep[e.g.,][]{2008ApJ...684..822B}.}

This argument can also be reversed for the highest values in the allowed range of the 1$\sigma$ confidence interval of the SMBH mass. Even for a maximally spinning black hole, SMBH masses of $M_\bullet>10^9$~\Msun would not lead to luminous emission from the disruption of a 2.5~\Msun~star. Or, in other words, ASASSN-15lh cannot be a TDE if the SMBH has $M_\bullet>10^9$~\Msun~(Fig.~\ref{fig:Hills}).

An alternative way to create luminous emission from a stellar disruption in such a massive galaxy could be a tight binary SMBH in which the two black holes have very different masses, and the lower mass secondary is sufficiently light to produce a bright TDE \citep{2017arXiv170504689C}. This scenario, however, would leave the total luminosity of ASASSN-15lh, which is approximately ten times higher compared to other TDEs, unexplained. In addition, SMBH binaries are rare and probably contribute only $\sim 3$~\% of the cosmic TDE rate \citep{2011ApJ...738L...8W}, and of those binary-induced disruptions, the fraction of TDEs caused by low-mass secondaries is only $\sim1$~\% \citep{2013PhDT.........2W}.

\subsection{Large-scale environment}

A comoving area of 220 kpc by 220 kpc around ASASSN-15lh is shown in Fig.~\ref{fig:fc} and contains a number of galaxies with a similar redshift. Within  a line-of-sight velocity of 200\,\kms, there are two early-type galaxies (EI and EII in Fig.~\ref{fig:fc}) at 40~kpc and 70~kpc projected distance, respectively, and a small satellite (20~kpc projected distance) to the north of the host that shows signs of star formation through the detection of the \ha~emission line (SFI). At a somewhat larger distance (100~kpc) and velocity offset (1800~\kms), two star-forming galaxies (SFII and SFIII) are connected through tidal tails and are therefore strongly interacting.

Remarkably, these six galaxies align along a specific direction (north-south). No other galaxies in the field of view with measured redshift are similarly close to ASASSN-15lh in velocity space, but these four galaxies (or six, when including the two more distant merging galaxies) constitute a significant galaxy overdensity. The ASSASN-15lh host is the most massive ($M_\star = 10.95^{+0.15}_{-0.11}$~\Msun) member observed, and possibly the node of a larger gravitationally bound system. Such an overdense environment is frequently observed around E+A galaxies as well \citep{2005MNRAS.357..937G}, and interactions are common. In fact, a major merger between two gas-rich spirals (SFII and SFIII) is even directly observed within this association.

The massive, early-type host of ASASSN-15lh is thus plausibly the result of a previous interaction, where the age of the youngest stellar population observed (1--2 Gyr, Sect.~\ref{sec:spop}) indicates the typical timescale of the merger. This scenario is broadly consistent with the LINER-like signatures in the central component \citep[e.g.,][]{2004ApJ...605..105C}, indicating the presence of a faint AGN,  early-type morphology, and other galaxy properties such as the lack of current star formation.

\subsection{Comparison to other TDE hosts}

It is useful to compare the ASASSN-15lh host to other well-observed galaxies that hosted more conventional, less extreme TDEs. The nearby TDE ASASSN-14li, for example, was hosted by a galaxy with comparable line ratios in the NLR, suggestive of a similar ionization and excitation process of the emission lines. In fact, the emission-line properties of the sample of TDE hosts studied by \citet{2017ApJ...835..176F} are in general in good agreement with the line properties observed here.

In contrast to ASASSN-14li, strong tidal tails are not observed in our case, neither in stellar light or in narrow emission lines. This might indicate that the ASASSN-15lh host is in a later stage after the galaxy-galaxy interaction with more time to relax into an undisturbed morphology, or simply that our spectroscopic observations are not deep enough to probe features of a past merger. In fact, most TDE hosts including ASASSN-14li have a relatively symmetric distribution of the stellar light  \citep{2017arXiv170701559L}, indicating that a previous galaxy interaction is not directly obvious in most cases.

The TDE rate seems to be significantly enhanced in E+A galaxies \citep{2014ApJ...793...38A, 2016ApJ...818L..21F}, which are thought to be observed few 100 Myrs after a starburst, likely induced through a galaxy merger \citep{2005MNRAS.357..937G}. In our case, the stellar population seems somewhat older (of order Gyr) but generally consistent with the distribution of ages and other properties in TDE host samples \citep{2017ApJ...835..176F, 2017arXiv170701559L}. It also displays LINER-like signatures that are also often present in E+A galaxies \citep{2006ApJ...646L..33Y}. This is again rather similar to the TDE hosts studied in \citet{2017ApJ...835..176F}, which in many cases are also LINERs.

These considerations thus indicate a common evolutionary path for the ASASSN-15lh host and other galaxies with TDEs. A galaxy-galaxy interaction leads to a starburst and possibly subsequent AGN activity, where the star formation has ceased to the present day, and the SMBH only accretes at a very low rate ($\lambda_{\mathrm{edd}}\sim10^{-5}-10^{-4}$) in our case). The nuclear starburst leads to a high stellar density, which in turn increases the TDE rate due to the short relaxation time for two-body interactions \citep{2016MNRAS.461..948M, 2016ApJ...825L..14S, 2017arXiv170702986G}. The evolutionary stage is set by the age of the youngest stellar component, and indicates a timescale of around a Gyr since the starburst, somewhat older than the average timescales of other TDE hosts. 

\section{Summary and conclusions}

We have presented here HST imaging and spatially resolved, medium-resolution spectroscopy from the ESO VLT instruments (X-Shooter and MUSE) of the environment of the luminous transient ASASSN-15lh. Based on these data, we reach the following conclusions:\\
\emph{(i)} The spectrum of the galaxy nucleus consists of three components: a stellar component with at least two different stellar populations (1-2 Gyr and 10 Gyr), the transient emission that is reasonably well described with a blackbody of $\sim$~13\,000~K (at 430 days after peak), and constant, narrow ($\sim$500~\kms) line emission from ionized gas.\\
\textit{(ii)} The line emission is related to the transitions of \ha, \hb, \nii, and \oiii, and splits up into three components that are separated spatially by 1.3~kpc and in velocity by 500~\kms. \\
\textit{(iii)} From their position in the BPT diagram, we show that the line ratios are consistent with LINER-like excitation, and we demonstrate that ionization by a weak AGN, and not star formation, is the likely origin of the observed emission lines.\\
\textit{(iv)} The central emission-line component is positionally coincident with the ASASSN-15lh transient, and we suggest that this is also the position of the central supermassive black hole in the host galaxy. The mass of the black hole derived from the stellar velocity dispersion $\sigma$ is $M_\bullet = 5.3_{-3.0}^{+8.0}\cdot10^{8} M_\sun$. We argue that the origin of the complex emission-line shape is best explained by kinematical effects (a rotating circumnuclear gas disk, or less-likely, outflows) in the AGN NLR.\\
\textit{(v)} The spatial association of the transient with a supermassive black hole, together with no detectable star formation, leads us to favor physical mechanisms for the transient that involve the supermassive black hole, and from those mechanisms specifically the tidal disruption of a star by a spinning SMBH.\\
\textit{(vi)} The observed physical properties of the ASASSN-15lh environment are in striking contrast to those of explosive phenomena related to the death of very massive stars such as hydrogen-poor SLSNe \citep[e.g.,][]{2014ApJ...787..138L, 2016arXiv161205978S} or long GRBs \citep[e.g.,][]{2015A&A...581A.125K, 2017MNRAS.467.1795L, 2017RSOS....470304S}. The environment is much more akin to those of TDEs, even though the post-starburst timescale of 1 to 2 Gyr observed here is slightly longer than the average of other TDE hosts \citep[e.g.,][]{2017ApJ...835..176F, 2017arXiv170701559L}, and the SMBH and stellar mass of the host is significantly higher.\\
This interpretation is consistent with the galaxy morphology and spectrum as a passively evolving galaxy, the location of the galaxy in an overdense larger scale environment, and the faint AGN emission in its center. All this suggests that the ASASSN-15lh host went through a very active phase of star formation and AGN activity roughly one or two Gyr in the past, possibly triggered by a galaxy-galaxy interaction.

\begin{acknowledgements}

We are very grateful to the referee for a timely and very constructive report, the language editor, as well as I.~Arcavi, L.~Christensen, J.~Greiner, P.~Schady, and L.~Izzo for helpful comments on the manuscript, which increased the clarity and quality of the manuscript. T.K. acknowledges support through the Sofja Kovalevskaja Award to P.~Schady from the Alexander von Humboldt Foundation of Germany. M.F. acknowledges the support of a Royal Society - Science Foundation Ireland University Research Fellowship. N.C.S. received financial support from NASA through Einstein Postdoctoral Fellowship Award Number PF5-160145, and thanks the Aspen Center for Physics for its hospitality during the completion of this work. DAK acknowledges support from the from the Spanish research project AYA 2014-58381-P and the Juan de la Cierva 
Incorporaci\'on fellowship IJCI-2015-26153. R.A. acknowledges support from the ERC Advanced Grant 695671 "QUENCH". We acknowledge the use of \texttt{NumPy} and \texttt{SciPy} \citep{Walt:2011:NAS:1957373.1957466} for computing and \texttt{matplotlib} \citep{Hunter:2007} for creating the plots in this manuscript. We thank ESO's Director's Discretionary Time Committee for allocating telescope time for this project, and the observing staff on Paranal for support in obtaining the MUSE and X-Shooter data.

\end{acknowledgements}

\bibliography{./bibtex/refs}

\end{document}